\documentclass[a4paper]{article}%

\usepackage{a4wide, authblk}
\usepackage{amsmath}
\usepackage{amsfonts}
\usepackage{amssymb}
\usepackage{amsthm}
\usepackage{bbm}
\usepackage{color}
\usepackage{graphicx}
\usepackage{mathrsfs}
\usepackage{enumitem}
\usepackage{xspace}
\usepackage{booktabs}
\usepackage[caption=false]{subfig}
\usepackage[english]{babel}
\usepackage{marvosym}
\usepackage{hyperref} 
\usepackage{multirow,tabularx}
\usepackage[utf8]{inputenc}

\newcommand{\beginsupplementnote}{%
	\setcounter{table}{0}
	\renewcommand{\tablename}{Supplementary Table}%
	\setcounter{figure}{0}
	\renewcommand{\figurename}{Supplementary Figure}
	\setcounter{section}{0}
	\renewcommand{\thesection}{Supplementary Note \arabic{section}}
	\renewcommand{\thesubsection}{\arabic{section}.\arabic{subsection}}
}

\newcommand{\sss}[1]{\scriptscriptstyle{#1}}
\newcommand{\mean}[1]{\left\langle #1 \right\rangle}

\title{Epidemic spreading on complex networks with community structures}
\author{Clara Stegehuis}
\author{Remco van der Hofstad}
\author{Johan S. H. van Leeuwaarden}
\affil{Eindhoven University of Technology}
\begin{document}
\maketitle
\begin{abstract}
Many real-world networks display a community structure. We study two random graph models that create a network with similar community structure as a given network. One model preserves the exact community structure of the original network, while the other model only preserves the set of communities and the vertex degrees.  These models show that community structure is an important determinant of the behavior of percolation processes on networks, such as information diffusion or virus spreading: the community structure can both \textit{enforce} as well as \textit{inhibit} diffusion processes. Our models further show that it is the mesoscopic set of communities that matters. The exact internal structures of communities barely influence the behavior of percolation processes across networks. This insensitivity is likely due to the relative denseness of the communities.
\end{abstract}

\maketitle
\footnotetext[1]{Eindhoven University of Technology, Department of Mathematics and Computer Science, P.O. Box 513, 5600 MB Eindhoven, The Netherlands.

}
\section*{Introduction}
Many complex systems across the sciences can be modeled as networks of vertices joined in pairs by edges.
Examples include the internet and the world-wide web, biological networks, food webs, the brain, neural networks, communication
and transport networks, and social networks.
This has spurred a tremendous interest in developing mathematical models that can capture universal network properties.
Moreover, with network data describing network topologies, properties derived from models can be tested against real-world networks.

The behavior of dynamic processes such as percolation or epidemic models on those networks are of significant interest, since for example they model the spreading of information or a virus across a network~\cite{dorogovtsev2008, boccaletti2006, pastor2014, barrat2008}. Understanding models for percolation may enhance insight in how an epidemic can be stopped by immunization, or how a message can go viral by choosing the right initial infectives. An important question is how the structure of the network affects the dynamics of the epidemic~\cite{newman2003book}. 
A vast amount of research focuses on scale-free networks that possess a power-law degree distribution \cite{clauset2009,hofstad2009, newman2010,newman2002b,vazquez2002}, so that the probability $p_k$ that a vertex has $k$ neighbors scales with $k$ as $p_k\sim c k^{-\tau}$ for some constant $c$ and characteristic exponent $\tau>1$. The power-law distribution leads to scale-free behavior such as short distances due to the likely presence of {\it hubs} or high-degree vertices. The characteristic exponent $\tau$ was also found to play a central role in various percolation processes~\cite{pastor2001, callaway2000, bhamidi2010b, bhamidi2012, cohen2001}. Other authors have focused on the influence of clustering on the spread of epidemics~\cite{gleeson2009, gleeson2010, serrano2006,serrano2006p2, trapman2007}.

Real-world networks, however, are not completely characterized by their microscopic and macroscopic properties. 
Many real-world networks display a community structure~\cite{girvan2002}, where groups of vertices are densely connected, while edges between different groups are more scarce. Since communities are small compared to the entire network, but seem to scale with the network size, they are typically of mesoscopic scale~\cite{fortunato2010, porter2009}. The problem of detecting the community structure of a network has received a lot of attention~\cite{fortunato2010, leskovec2010}. The exact way in which communities influence the properties of a network is a different problem. For example, the community structure of a network influences the way a cooperation process behaves on real-world networks~\cite{lozano2008}, and using community structure improves the prediction of which messages will go viral across a network~\cite{weng2013}.
Several stylized random graph models with a community structure have shown that communities influence the process of an epidemic across a network~\cite{salathe2010, liu2005, ball2010, bonaccorsi2014, gleeson2008,huang2007, yan2007, wu2008}, but the extent to which community structure affects epidemics on real-world networks is largely unexplained.
Our main goal is to enhance our understanding of the intricate relation between community structures and the spread of epidemics, and in particular to identify the properties of community structures that have the largest influence.

We study two random graph models that generate networks with a similar community structure as any given network. We find that these models capture the behavior of epidemics or percolation on real-world networks accurately, and that the mesoscopic community structure is vital for understanding epidemic spreading.
We find that the sets of communities are of crucial importance, while quite surprisingly, the precise structure of the intra-community connections hardly influences the percolation process. Furthermore, we find that community structure can both enforce as well as inhibit percolation.

\section*{Models}
\begin{figure*}[t]
\centering
\includegraphics[ width=0.80\textwidth]{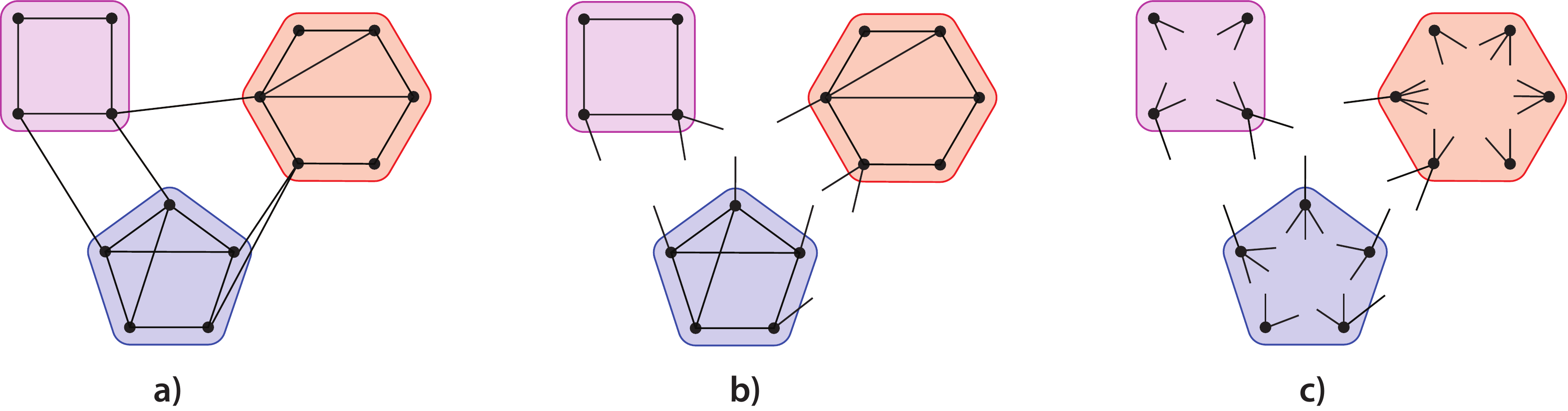}
\caption{\textbf{HCM and HCM* illustrated.} a) A network with 3 communities. b) HCM randomizes the edges between different communities. c) HCM* also randomizes the edges inside the communities.}
\label{fig:hcm}
\end{figure*}

We now introduce two random graph models in detail. For a given real-world network, both models randomize the edges of the network, while keeping large parts of the community imprint. Suppose that we are given the set of communities of a particular real-world network. Then the first model, the hierarchical configuration model (HCM), keeps all edges inside the communities~\cite{stegehuis2015, hofstad2015}, while rewiring the inter-community edges. Indeed, all inter-community edges are replaced by two half-edges, one at each end of an inter-community edge. Then, one by one, these half-edges are paired at random. Thus, in HCM, the precise community structure of the network is the same as in the original network, but the inter-community connections are random. The second model (HCM*), introduced as the modular random graph in~\cite{sah2014}, replaces both the inter-community edges and the intra-community edges by pairs of half-edges. Then again, the half-edges are paired at random. An additional constraint is that all inter-community half-edges must be paired to one another, and all half-edges corresponding to the same community must be paired to one another (see Figure~\ref{fig:hcm} and Supplementary Note 3). Thus, a network generated by HCM* is completely random, except for the set of communities and the degree distributions inside and outside the communities. 

HCM and HCM* are extensions of the configuration model (CM), a random graph with a given degree distribution. The CM has received enormous attention in the network literature, due to the combination of its simplicity and its flexibility in choosing an appropriate degree structure~\cite{molloy1995, newman2001}. CM only preserves the microscopic degree distribution of the real-world networks, while HCM* also preserves the mesoscopic community structure. HCM instead, preserves the entire community structure. 
Supplementary Table 3 shows that indeed most of the community structures of the original networks and the networks generated by HCM and HCM* are similar.
Therefore, if we sort the random graph models in decreasing randomness, we first have CM, then HCM*, and then HCM. 
When comparing the behavior of an epidemic process on these random graphs to the original network, we see how much of the behavior of epidemics on real-world networks can be explained by its degree distribution (CM), its rough community structure (HCM*), and by the exact community shapes (HCM). The aim of this paper is to investigate to which extent microscopic and mesoscopic network properties determine the spread of epidemics.

The fixed community shapes combined with the randomized inter-community connections make HCM analytically tractable~\cite{stegehuis2015}. However, keeping all intra-community edges fixed makes HCM prone to overfitting. HCM* does not have this problem and is more suitable to generate a random network with a community structure, since all edges within communities are randomized. Randomizing the intra-community edges makes HCM* harder to analyze analytically than HCM. Some analytical results of HCM, however, can be extended to results of HCM* (Supplementary Note 3). 

\section*{Results}

We analyze six different real-world networks: the internet on the Autonomous Systems level~\cite{snap}, an email network of the company Enron~\cite{klimt2004, snap}, the PGP web of trust~\cite{boguna2004}, a collaboration network in High energy physics, extracted from the arXiv~\cite{snap}, a \textsc{Facebook} friendship network~\cite{viswanath2009} and an interaction network between proteins in yeast~\cite{bu2003}. Table~\ref{tab:data} shows several statistics of these data sets and their community structures.
 We extract the communities of these networks with the Infomap community detection algorithm~\cite{rosvall2008}, and use these communities as input for the HCM and HCM* model, to create networks with a similar community structure as the original networks. 
 Table~\ref{tab:data} shows that the communities are of mesoscopic size: while the communities are small compared to the entire network, and have a small expected size, all networks still contain a few large communities.
\begin{table}[t]
  \centering
    \begin{tabular}{lrrrrrr}
    \textbf{} & $N$     &  $\mean{s}$ &$s_{\max}$  & $\delta_{\text{netw}} $ &$\delta_{\text{com}}$  &$\delta_{\text{com}}^w$   \\
    \midrule
    \textbf{AS} & 11,174 &  21  &910   & 3.75 $ \cdot 10^{-4}$&  0.38  & 0.10\\
    \textbf{Enron} & 36,692&  15 & 1,722   &  2.73 $ \cdot 10^{-4}$&0.73  & 0.22\\
    \textbf{HEP} & 9,877  &  10 &181  &  5.33 $ \cdot 10^{-4}$& 0.59  & 0.32 \\
    \textbf{PGP} & 10,680 & 12 &160   & 4.26 $ \cdot 10^{-4}$& 0.41 & 0.24\\
    \textbf{FB} & 63,731 &  29 &2,247  & 4.02 $ \cdot 10^{-4}$& 0.41 & 0.14 \\
    \textbf{yeast} & 2,361  & 9 &97   &  2.57 $ \cdot 10^{-3}$& 0.55   & 0.25\\
    \bottomrule
    \end{tabular}%
      \caption{Statistics of the data sets. $N$ is the number of vertices in the network, $\mean{s}$ the average community size, $s_{\max}$ the maximal community size. The denseness of the network  $\delta_{\text{netw}}$ is defined as the number of edges divided by the number of edges in a complete graph of the same size.  $\delta_{\text{com}}$ equals the average denseness of the communities, and $\delta_{\text{com}}^{w}$  the average denseness of the communities weighted by their sizes (See Supplementary Note 1 for more information about these statistics).}
  \label{tab:data}%
\end{table}%

An important property of a network is its connectedness, expressed by the fraction of vertices in the largest component. For HCM, the size of the largest component can be derived analytically (Supplementary Note 3). This size is independent of the precise community shapes, and therefore is the same for HCM and HCM*, as long as the communities of HCM* remain connected. 
Supplementary Note 3.3 shows that most HCM* communities indeed remain connected.
The size of the largest component of real-world networks can be well predicted using the analytical estimates of HCM, which only uses the joint distribution of community sizes and the number of edges going out of the communities (Table~\ref{tab:S}). These estimates yield a considerable improvement compared to CM, which is generally a few percent off.

\begin{table}[t]
  \centering
    \begin{tabular}{lrrrr}
          & $S$ (data) & $S$ (HCM) &  $S$ (HCM*)& $S$ (CM) \\
    \midrule
    \textbf{AS} & 1.000 & 1.000 & 1.000& 0.960 \\
    \textbf{Enron} & 0.918 & 0.918 &0.918 & 0.990 \\
    \textbf{HEP} & 0.875 & 0.875 & 0.875 &0.990 \\
    \textbf{PGP} & 1.000 & 1.000 &1.000 & 0.960 \\
    \textbf{FB} & 0.995 & 0.995 &  0.995 & 0.999 \\
    \textbf{yeast} & 0.941 & 0.941 & 0.941 & 0.948 \\
    \bottomrule
    \end{tabular}%
      \caption{The size $S$ of the giant component in the data sets compared to the analytical estimates of HCM and CM.}
  \label{tab:S}%
\end{table}%

\begin{figure*}[t]
	\centering
		\includegraphics[ width=0.65\textwidth]{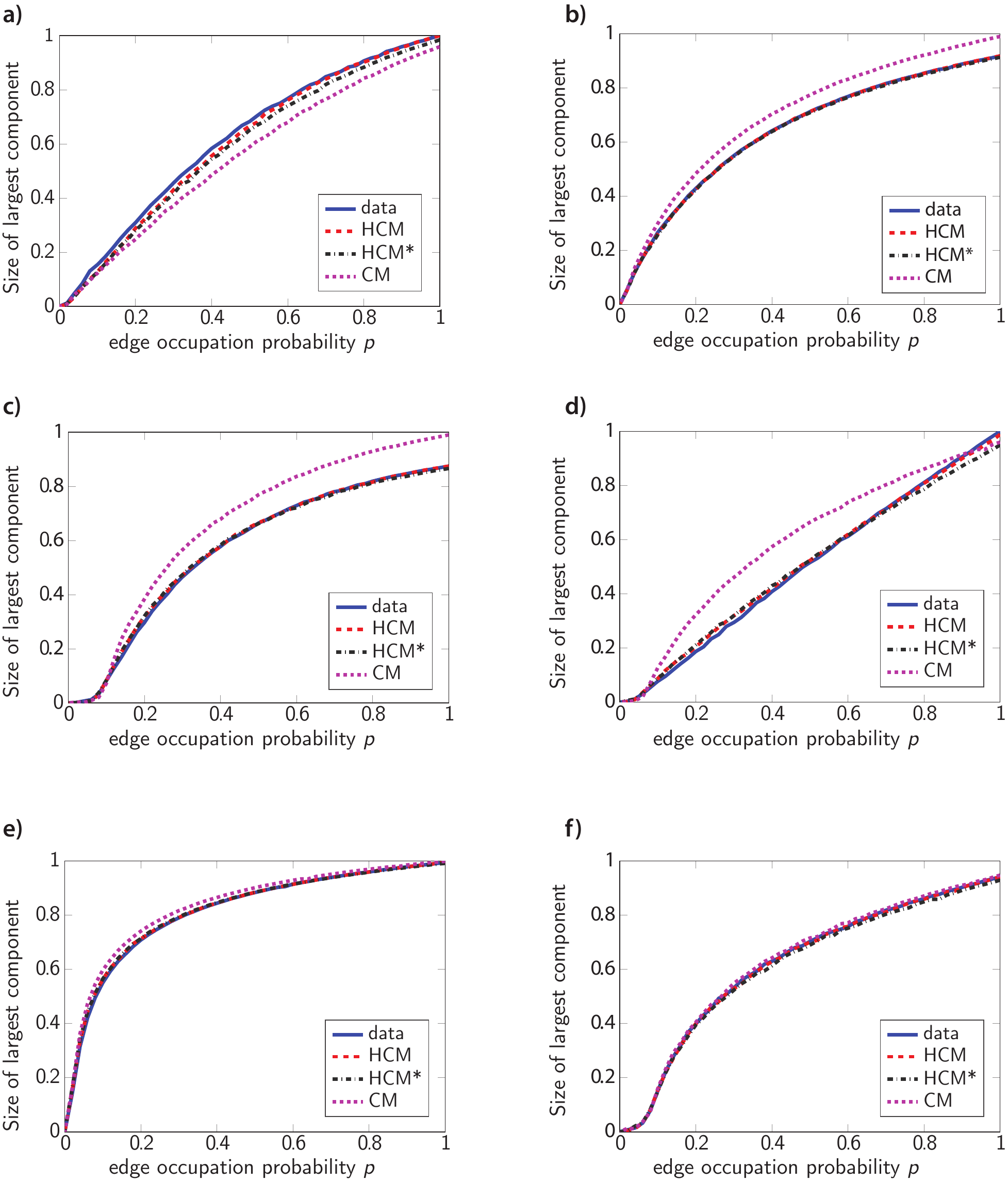}
	\caption{\textbf{HCM, HCM* and CM under bond percolation compared to real-world networks.} a) Autonomous Systems network b) \textsc{Enron} email network c) Collaboration network in High energy physics d) PGP network e) \textsc{Facebook} friendship network f) yeast network. Independently, each edge is deleted with probability $1-p$. The size of the largest component after deleting the edges is the average of 500 generated graphs.} 
	\label{fig:perc}	
\end{figure*}

The long-term properties of an epidemic outbreak can be mapped into a suitable bond percolation
problem. In this framework, the probability $p$ that a link exists is related to the probability
of transmission of the disease from an infected vertex to a connected susceptible vertex. The latter corresponds to removing edges in a network with probability $1-p$ and keeping the edges with probability $p$ independently across edges (other types of epidemics are discussed in Supplementary Note 4).
A quantity of interest is the size of the largest component as a function of $p$, which can be described analytically for HCM~\cite{stegehuis2015}. However, this size depends on the community shapes, and therefore bond percolation on HCM does not necessarily give the same results as percolation on HCM*. Inspired by the insensitivity of the giant component to the exact community shapes, we establish whether the community shapes significantly influence the size of the giant percolating cluster by simulation, by showing how bond percolation affects the connectivity of the original networks, compared to CM, HCM and HCM* (Figure~\ref{fig:perc}). 

We see that the behavior of the real-world networks under bond percolation is captured accurately by both HCM and HCM*, in contrast to CM. In Supplementary Figures~1\--5, we see that HCM and HCM* also perform well for other types of percolation processes and an SIR epidemic.
These results reveal and confirm the key role of the mesoscopic community structure in percolation processes. Furthermore, the fact that the predictions of HCM and HCM* are \textit{both} close to the behavior of the original network under percolation indicates that the shapes of the communities only have minor influence on the percolation process. 
The surprising finding that the exact internal community structure barely influences the epidemic processes may be explained by the denseness of the communities. Table~\ref{tab:data} shows that the communities are very dense compared to the entire network. Since community detection algorithms look for dense subsets in large complex networks, applying HCM or HCM* to real-world networks typically yields sets of dense communities. The Autonomous Systems network has communities that are much less dense than in most other networks~\cite{lancichinetti2010}, but even in that network the communities are much denser than the entire network. Therefore, in the case of bond percolation for example, the communities of mesoscopic size are supercritical, and the communities will be almost connected after percolation.
Thus, an epidemic entering a community of mesoscopic size will reach most other community members. It is more difficult for the epidemic to reach other communities, which makes the inter-community edges the important factor for the spread of an epidemic. When generating a HCM* network, the communities stay of the same denseness, and therefore it is still relatively easy for the epidemic to spread inside the communities, regardless of their exact shapes.

The only process where HCM and HCM* are not always close to the process on the original graph, is a targeted attack (Supplementary Figure~2), even though both models still outperform CM. Furthermore, some networks show a difference between the predictions of HCM and HCM*. Therefore, the exact community structures may have some influence on a targeted attack on a real-world network.
Another interesting observation is that where most networks are highly sensitive to a targeted attack, the \textsc{Facebook} network has a community structure that makes it more resistant against a targeted attack than a configuration model. 
This particular feature of the \textsc{Facebook} network can be explained by the fact that in the \textsc{Facebook} network, most vertices of high degree are in the same community. Therefore, deleting high degree vertices has a smaller effect than in a corresponding CM model. 

The results of the yeast network show that in some situations CM performs equally well as HCM or HCM*. Thus, in some cases the mesoscopic properties of a network do not influence percolation processes. In the case of the yeast network, this can be explained by its almost tree-like structure; there is no noticeable community structure. Thus, by adding the community structure in HCM or HCM*, no structural information is added. 
This suggests that CM, HCM and HCM* combined can also show whether the community structure given by a community detection algorithm is meaningful. When the behavior of various processes on CM, HCM and HCM* are similar, this may imply that there is no real community structure in the network.

The \textsc{Enron}, High Energy Physics and PGP networks have communities that inhibit percolation or an SIR epidemic compared to a configuration model with the same degree distribution. This is similar to the observation that communities can act as traps for an epidemic process across a network~\cite{onnela2007}.
In contrast, the communities in the Autonomous Systems graph enforce the percolation process, which may be attributed to its star-like community structure. Since HCM* preserves the degrees of the vertices inside their own community, HCM* creates a graph that captures this star-like structure. 
An important conclusion is that these findings confirm that both HCM and HCM* are realistic models for real-world networks.

Where~\cite{orsini2015} creates a reshuffling of a given network using several microscopic properties of every vertex, HCM and HCM* use mesoscopic properties instead. An advantage of using HCM or HCM* is that both models are easy to generate. 
Since HCM* is more random than HCM, it is a better choice for generating a random network. Note that in HCM*, the rewiring of intra-community edges makes the community structure a uniform simple graph with the prescribed degrees.
Specifically, if the interest is to generate a random graph such that percolation on that graph behaves in a similar way as in the original network, then our results show that HCM* is a suitable choice. However, HCM* does not capture the microscopic properties of the original network as effectively as HCM. HCM*, for example, does not generate networks with similar clustering as in the original network~\cite{sah2014}.
Therefore, when the goal is to create a network with similar clustering as the original network, using HCM* may be less suitable. Indeed Table~\ref{tab:clust} shows that in most cases HCM generates a network with a clustering coefficient that is closer to the value of the original network. An exception is the Autonomous Systems network, where HCM* is closer to the real value of the clustering. An explanation for this is that the communities in the Autonomous Systems network have virtually no clustering; all clustering is between different communities. HCM also has no clustering inside the communities, but the pairing between different communities destroys the clustering between different communities, and therefore HCM creates a network with a lower clustering coefficient. HCM* also destroys the clustering between different communities, but by rewiring the edges inside communities, creates some clustering inside the communities. Therefore, the value of the clustering of HCM* is closer to the value of the original network than the one of HCM. 

The fact that HCM* does not capture the clustering coefficient and the assortativity (See Supplementary Notes 3.1) well, but does capture the spread of an epidemic across a network, again confirms that the mesoscopic properties are of vital importance for the spread of an epidemic across a network. Even though microscopic features such as clustering are destroyed in HCM*, the mesoscopic properties are sufficient to know how an epidemic spreads, making HCM* a suitable random graph model when considering the mesoscopic structure of networks.

\begin{table}[tbp]
  \centering
    \begin{tabular}{lrrrr}
    \textbf{} & \textbf{data} & \textbf{HCM} & \textbf{HCM*} & \textbf{CM} \\
    \midrule
    \textbf{AS} & 0.30  & 0.16  & 0.20  & 0.09 \\
    \textbf{Enron} & 0.50  & 0.35  & 0.22  & 0.03 \\
    \textbf{HEP} & 0.47  & 0.40  & 0.24  & 0.00 \\
    \textbf{PGP} & 0.26  & 0.24  & 0.19  & 0.00 \\
    \textbf{FB} & 0.22  & 0.15  & 0.08  & 0.00 \\
    \textbf{yeast} & 0.13  & 0.12  & 0.12  & 0.01 \\
    \bottomrule
    \end{tabular}%
      \caption{Average clustering for the original data set, HCM, HCM* and CM. The presented values are averages of 100 generated graphs.}
  \label{tab:clust}%
\end{table}%

\section*{Conclusion}
Community structures in real-world networks have a profound impact on percolation or epidemic spreading, which is central to our understanding of dynamical processes in complex networks. The theoretical analysis of epidemic spreading in heterogeneous networks with community structure
requires the development of novel analytical frameworks. We have introduced the hierarchical configuration model (HCM) to describe such networks. Both HCM and its randomized counterpart HCM* turn out be highly suitable to capture epidemic spreading on real-world networks. We have shown this by mapping the models to various real-world networks, and by investigating a range of epidemic processes including bond percolation, bootstrap percolation and an SIR epidemic. Our experiments show that while it is essential to take the community structure into account, the precise internal structure of communities is far less important for describing an epidemic outbreak. This insensitivity is likely due to the relative denseness of the communities. When communities are sparse, their internal structures are expected to have a more decisive effect on epidemic spreading. The HCM and HCM* models can easily be extended to include overlapping communities, by considering an auxiliary graph. It would be interesting to see whether including overlapping communities further improves the description of percolation across complex networks.

\subsection*{Acknowledgement}
This work is supported by NWO TOP grant 613.001.451 and by the NWO Gravitation Networks grant 024.002.003.
The work of RvdH is further supported by the NWO VICI grant 639.033.806.  The work of JvL is further supported by an NWO TOP-GO grant and by an ERC Starting Grant.

\nocite{newman2004, newman2002,litvak2013,lancichinetti2008,milo2003}

\bibliographystyle{unsrt}
\bibliography{../references}

\begin{thebibliography}{10}

\bibitem{dorogovtsev2008}
Sergey~N Dorogovtsev, Alexander~V Goltsev, and Jos{\'e}~FF Mendes.
\newblock Critical phenomena in complex networks.
\newblock {\em Reviews of Modern Physics}, 80(4):1275, 2008.

\bibitem{boccaletti2006}
Stefano Boccaletti, Vito Latora, Yamir Moreno, Martin Chavez, and D-U Hwang.
\newblock Complex networks: Structure and dynamics.
\newblock {\em Physics Reports}, 424(4):175--308, 2006.

\bibitem{pastor2014}
Romualdo Pastor-Satorras, Claudio Castellano, Piet Van~Mieghem, and Alessandro
  Vespignani.
\newblock Epidemic processes in complex networks.
\newblock {\em Rev. Mod. Phys.}, 87:925--979, Aug 2015.

\bibitem{barrat2008}
Alain Barrat, Marc Barthelemy, and Alessandro Vespignani.
\newblock {\em Dynamical processes on complex networks}.
\newblock Cambridge University Press, 2008.

\bibitem{newman2003book}
Mark E.~J. Newman.
\newblock The structure and function of complex networks.
\newblock {\em SIAM Review}, 45(2):167--256, 2003.

\bibitem{clauset2009}
Aaron Clauset, Cosma~Rohilla Shalizi, and M.~E.~J. Newman.
\newblock Power-law distributions in empirical data.
\newblock {\em SIAM Review}, 51(4):661--703, 2009.

\bibitem{hofstad2009}
Remco {\swap{Hofstad}{van der }}.
\newblock Random {G}raphs and {C}omplex {N}etworks {Vol. I}.
\newblock {\em To appear with Cambridge University Press}, 2016.

\bibitem{newman2010}
Mark E.~J. Newman.
\newblock {\em Networks: {An} introduction}.
\newblock Oxford University Press, 2010.

\bibitem{newman2002b}
M.~E.~J. Newman, Stephanie Forrest, and Justin Balthrop.
\newblock Email networks and the spread of computer viruses.
\newblock {\em Phys. Rev. E}, 66:035101, Sep 2002.

\bibitem{vazquez2002}
Alexei V{\'a}zquez, Romualdo Pastor-Satorras, and Alessandro Vespignani.
\newblock Large-scale topological and dynamical properties of the internet.
\newblock {\em Phys. Rev. E}, 65(6):066130, 2002.

\bibitem{pastor2001}
Romualdo Pastor-Satorras and Alessandro Vespignani.
\newblock Epidemic spreading in scale-free networks.
\newblock {\em Phys. Rev. Lett.}, 86(14):3200, 2001.

\bibitem{callaway2000}
Duncan~S Callaway, Mark E.~J. Newman, Steven~H Strogatz, and Duncan~J Watts.
\newblock Network robustness and fragility: Percolation on random graphs.
\newblock {\em Phys. Rev. Lett.}, 85(25):5468, 2000.

\bibitem{bhamidi2010b}
Shankar Bhamidi, Remco van~der Hofstad, and Johan van Leeuwaarden.
\newblock Scaling limits for critical inhomogeneous random graphs with finite
  third moments.
\newblock {\em Electron. J. Probab.}, 15:no. 54, 1682--1702, 2010.

\bibitem{bhamidi2012}
Shankar Bhamidi, Remco van~der Hofstad, and Johan S.~H. van Leeuwaarden.
\newblock Novel scaling limits for critical inhomogeneous random graphs.
\newblock {\em Ann. Probab.}, (6):2299--2361, 11.

\bibitem{gleeson2009}
James~P Gleeson.
\newblock Bond percolation on a class of clustered random networks.
\newblock {\em Physical Review E}, 80(3):036107, 2009.

\bibitem{gleeson2010}
James~P Gleeson, Sergey Melnik, and Adam Hackett.
\newblock How clustering affects the bond percolation threshold in complex
  networks.
\newblock {\em Physical Review E}, 81(6):066114, 2010.

\bibitem{serrano2006}
M~{\'A}ngeles Serrano and Mari{\'a}n Bogun{\'a}.
\newblock Percolation and epidemic thresholds in clustered networks.
\newblock {\em Physical Review Letters}, 97(8):088701, 2006.

\bibitem{serrano2006p2}
M~{\'A}ngeles Serrano and Mari{\'a}n Bogun{\'a}.
\newblock Clustering in complex networks. {II.} percolation properties.
\newblock {\em Physical Review E}, 74(5):056115, 2006.

\bibitem{trapman2007}
Pieter Trapman.
\newblock On analytical approaches to epidemics on networks.
\newblock {\em Theoretical Population Biology}, 71(2):160 -- 173, 2007.

\bibitem{girvan2002}
Michelle Girvan and Mark~EJ Newman.
\newblock Community structure in social and biological networks.
\newblock {\em Proceedings of the National Academy of Sciences},
  99(12):7821--7826, 2002.

\bibitem{fortunato2010}
Santo Fortunato.
\newblock Community detection in graphs.
\newblock {\em Physics Reports}, 486(3):75--174, 2010.

\bibitem{porter2009}
Mason~A Porter, Jukka-Pekka Onnela, and Peter~J Mucha.
\newblock Communities in networks.
\newblock {\em Notices of the AMS}, 56(9):1082--1097, 2009.

\bibitem{leskovec2010}
Jure Leskovec, Kevin~J Lang, and Michael Mahoney.
\newblock Empirical comparison of algorithms for network community detection.
\newblock In {\em Proceedings of the 19th international conference on World
  wide web}, pages 631--640. ACM, 2010.

\bibitem{lozano2008}
Sergi Lozano, Alex Arenas, and Angel S{\'a}nchez.
\newblock Mesoscopic structure conditions the emergence of cooperation on
  social networks.
\newblock {\em PLoS One}, 3(4):e1892, 2008.

\bibitem{weng2013}
Lilian Weng, Filippo Menczer, and Yong-Yeol Ahn.
\newblock Virality prediction and community structure in social networks.
\newblock {\em Scientific Reports}, 3, 2013.

\bibitem{salathe2010}
Marcel Salath{\'e} and James~H Jones.
\newblock Dynamics and control of diseases in networks with community
  structure.
\newblock {\em PLoS Comput Biol}, 6(4):e1000736, 2010.

\bibitem{liu2005}
Zonghua Liu and Bambi Hu.
\newblock Epidemic spreading in community networks.
\newblock {\em EPL (Europhysics Letters)}, 72(2):315, 2005.

\bibitem{ball2010}
Frank Ball, David Sirl, and Pieter Trapman.
\newblock Analysis of a stochastic {SIR} epidemic on a random network
  incorporating household structure.
\newblock {\em Mathematical Biosciences}, 224(2):53 -- 73, 2010.

\bibitem{bonaccorsi2014}
Stefano Bonaccorsi, Stefania Ottaviano, Francesco De~Pellegrini, Annalisa
  Socievole, and Piet Van~Mieghem.
\newblock Epidemic outbreaks in two-scale community networks.
\newblock {\em Phys. Rev. E}, 90(1):012810, 2014.

\bibitem{gleeson2008}
James~P Gleeson.
\newblock Cascades on correlated and modular random networks.
\newblock {\em Phys. Rev. E}, 77(4):046117, 2008.

\bibitem{huang2007}
Wei Huang and Chunguang Li.
\newblock Epidemic spreading in scale-free networks with community structure.
\newblock {\em Journal of Statistical Mechanics: Theory and Experiment},
  2007(01):P01014, 2007.

\bibitem{yan2007}
Gang Yan, Zhong-Qian Fu, Jie Ren, and Wen-Xu Wang.
\newblock Collective synchronization induced by epidemic dynamics on complex
  networks with communities.
\newblock {\em Phys. Rev. E}, 75(1):016108, 2007.

\bibitem{wu2008}
Xiaoyan Wu and Zonghua Liu.
\newblock How community structure influences epidemic spread in social
  networks.
\newblock {\em Physica A: Statistical Mechanics and its Applications},
  387(2):623--630, 2008.

\bibitem{stegehuis2015}
Clara Stegehuis, Remco {\swap{Hofstad}{van der }}, and Johan S.~H.
  {\swap{Leeuwaarden}{van }}.
\newblock Power-law relations in random networks with communities.
\newblock {\em Phys. Rev. E}, 94:012302, Jul 2016.

\bibitem{hofstad2015}
Remco {\swap{Hofstad}{van der }}, Johan S.~H. {\swap{Leeuwaarden}{van }}, and
  Clara Stegehuis.
\newblock Hierarchical configuration model.
\newblock {\em arXiv:1512.08397}, 2015.

\bibitem{sah2014}
Pratha Sah, Lisa~O Singh, Aaron Clauset, and Shweta Bansal.
\newblock Exploring community structure in biological networks with random
  graphs.
\newblock {\em BMC Bioinformatics}, 15(1):220, 2014.

\bibitem{molloy1995}
Michael Molloy and Bruce Reed.
\newblock A critical point for random graphs with a given degree sequence.
\newblock {\em Random Structures \& Algorithms}, 6(2-3):161--180, 1995.

\bibitem{newman2001}
Mark E.~J. Newman, Steven~H Strogatz, and Duncan~J Watts.
\newblock Random graphs with arbitrary degree distributions and their
  applications.
\newblock {\em Phys. Rev. E}, 64(2):026118, 2001.

\bibitem{snap}
Jure Leskovec and Andrej Krevl.
\newblock {SNAP Datasets}: {Stanford} large network dataset collection.
\newblock \url{http://snap.stanford.edu/data}, June 2014.
\newblock Date of access: 12/12/2015.

\bibitem{klimt2004}
Bryan Klimt and Yiming Yang.
\newblock Introducing the {E}nron {C}orpus.
\newblock In {\em CEAS}, 2004.

\bibitem{boguna2004}
Mari\'an Bogu\~n\'a, Romualdo Pastor-Satorras, Albert D\'{\i}az-Guilera, and
  Alex Arenas.
\newblock Models of social networks based on social distance attachment.
\newblock {\em Phys. Rev. E}, 70:056122, Nov 2004.

\bibitem{viswanath2009}
Bimal Viswanath, Alan Mislove, Meeyoung Cha, and Krishna~P Gummadi.
\newblock On the evolution of user interaction in facebook.
\newblock In {\em Proceedings of the 2nd ACM workshop on Online social
  networks}, pages 37--42. ACM, 2009.

\bibitem{bu2003}
Dongbo Bu, Yi~Zhao, Lun Cai, Hong Xue, Xiaopeng Zhu, Hongchao Lu, Jingfen
  Zhang, Shiwei Sun, Lunjiang Ling, Nan Zhang, et~al.
\newblock Topological structure analysis of the protein--protein interaction
  network in budding yeast.
\newblock {\em Nucleic Acids Research}, 31(9):2443--2450, 2003.

\bibitem{rosvall2008}
Martin Rosvall and Carl~T. Bergstrom.
\newblock Maps of random walks on complex networks reveal community structure.
\newblock {\em Proceedings of the National Academy of Sciences},
  105(4th):1118--1123, 2008.

\bibitem{lancichinetti2010}
Andrea Lancichinetti, Mikko Kivelä, Jari Saramäki, and Santo Fortunato.
\newblock Characterizing the community structure of complex networks.
\newblock {\em PLoS ONE}, 5, 08 2010.

\bibitem{onnela2007}
J-P Onnela, Jari Saram{\"a}ki, Jorkki Hyv{\"o}nen, Gy{\"o}rgy Szab{\'o}, David
  Lazer, Kimmo Kaski, J{\'a}nos Kert{\'e}sz, and A-L Barab{\'a}si.
\newblock Structure and tie strengths in mobile communication networks.
\newblock {\em Proceedings of the National Academy of Sciences},
  104(18):7332--7336, 2007.

\bibitem{orsini2015}
Chiara Orsini, Marija~M Dankulov, Pol Colomer-de Sim{\'o}n, Almerima Jamakovic,
  Priya Mahadevan, Amin Vahdat, Kevin~E Bassler, Zolt{\'a}n Toroczkai,
  Mari{\'a}n Bogu{\~n}{\'a}, Guido Caldarelli, et~al.
\newblock Quantifying randomness in real networks.
\newblock {\em Nature Communications}, 6, 2015.

\bibitem{newman2004}
M.~E.~J. Newman and M.~Girvan.
\newblock Finding and evaluating community structure in networks.
\newblock {\em Phys. Rev. E}, 69:026113, Feb 2004.

\bibitem{newman2002}
M.~E.~J. Newman.
\newblock Mixing patterns in networks.
\newblock {\em Phys. Rev. E}, 67:026126, Feb 2003.

\bibitem{litvak2013}
Nelly Litvak and Remco {\swap{Hofstad}{van der }}.
\newblock Uncovering disassortativity in large scale-free networks.
\newblock {\em Phys. Rev. E}, 87:022801, Feb 2013.

\bibitem{lancichinetti2008}
Andrea Lancichinetti, Santo Fortunato, and Filippo Radicchi.
\newblock Benchmark graphs for testing community detection algorithms.
\newblock {\em Phys. Rev. E}, 78(4):046110, 2008.

\bibitem{milo2003}
R~Milo, N~Kashtan, S~Itzkovitz, M.~E.~J. Newman, and U~Alon.
\newblock On the uniform generation of random graphs with prescribed degree
  sequences.
\newblock {\em arXiv preprint cond-mat/0312028}, 2003.

\end{thebibliography}

\newpage

\beginsupplementnote
\section*{Supplementary Notes}
\section{Network properties}\label{sec:info}
Tables~\ref{tab:data},~\ref{tab:S} and~\ref{tab:clust} give several statistics of the networks that we have considered. We now explain these characteristics in more detail.

\subsection{Graphs}
A graph $G=(V,E)$ consists of a set of vertices $V$, and a set of edges $E$. In this paper, all graphs are undirected. The edge set $E$ consists of pairs of vertices that are linked to one another, so that if $\{v_1,v_2\}\in E$, then $v_1$ and $v_2$ have an edge between them. The degree of vertex $i$ is denoted by $d_i$, and represents the number of edges that are adjacent to vertex $i$. Let $N$ denote the number of vertices in the graph, and $N_k$ the number of vertices in the graph of degree $k$. Then the degree distribution of the graph is given by $p_k=N_k/N$, for $k=0,1,\dots$.

\subsection{Clustering}
The clustering coefficient $C_i$ of a vertex $i$ is defined as the number of triangles that $i$ is part of, $T_i$, divided by the number of pairs of neighbors of vertex $i$, so that 
\begin{equation}
C_i=\frac{2 T_i}{d_i(d_i-1)}.
\end{equation}
This can be interpreted as the fraction of neighbors of $v$ that are also neighbors of one another. Given the degree $d_i$ of vertex $i$, the number of pairs of neighbors of $i$ equals $d_i(d_i-1)/2$. Then the average clustering coefficient $C$ is defined as the average of the clustering coefficient of all vertices,
\begin{equation}
C=\frac{1}{N}\sum_{i=1}^NC_i.
\end{equation}

\subsection{Modularity}
The modularity of a network is a measure of how well the network can be divided into communities. Consider a partition $P$ of the $N$ vertices into communities. 
The modularity $M(P)$ of a partition $P$ equals~\cite{newman2004}
\begin{equation}
M(P)=\sum_{c\in P}\frac{L_c}{L}-\left(\frac{d_c}{2L}\right)^2,
\end{equation}
where $L$ is the number of edges of the network, $L_c$ is the number of edges inside community $c$, and $d_c$ is the sum of all degrees of vertices in community $c$. This is a measure of how many edges are inside the community, minus how many edges would be expected if the vertices were connected at random. Therefore, a higher modularity implies many edges inside communities. 

\subsection{Assortativity}
The assortativity of a graph $G=(V,E)$ can be interpreted as the correlation between the degrees at the end of a randomly chosen edge~\cite{newman2002} and is given by
\begin{equation}\label{eq:assort}
r(G)=\frac{2\sum_{\{i,j\}\in E}d_id_j-\frac{1}{2L}\left(\sum_i d_i^2\right)^2}{\sum_id_i^3-\frac{1}{2L}\left(\sum_i d_i^2\right)^2}.
\end{equation}
Positive assortativity indicates that vertices of high degree are connected to other vertices of high degree, and negative assortativity indicates that high degree vertices are typically connected to vertices of low degree. Assortativity is a frequently used network statistic, despite its dependence on the network size~\cite{litvak2013}.

\subsection{Size of giant component}
A network may consist of several connected components, $\mathscr{C}_1,\mathscr{C}_2,\dots,\mathscr{C}_k$. The proportion of vertices in the giant component $S$ is then defined as the fraction of vertices in the largest connected component: $S=|\mathscr{C}_{\text{max}}|/N$, where $\mathscr{C}_{\text{max}}$ is the largest connected component in the network, i.e., $|\mathscr{C}_{\text{max}}|=\max_{i}|\mathscr{C}_i|$, where $|\mathscr{C}|$ denotes the size of cluster $\mathscr{C}$. If the maximal component is not unique, we break ties in an arbitrary way.

\subsection{Graph distances}
The graph distance between two vertices $u$ and $v$ is defined as the minimal number of edges in a path that links $u$ and $v$. Supplementary Figure~\ref{fig:dist} presents the graph distances for the different data sets. In some instances, HCM and HCM* capture the graph distances better than CM. However, for example for the yeast network, the distances in CM are already close to the distances in the original data set. 

\section{Community detection}\label{sec:detect}
The HCM and HCM* models use as input the community structure of a network. Several algorithms to detect this community structure are available. In this paper, we used the Infomap community detection algorithm~\cite{rosvall2008}. This community detection algorithm uses a random walk perspective to detect the communities. It has a computational complexity of $O(N\log(N))$ for a network of $N$ vertices, making it applicable to detect community structures in large networks. Furthermore, the algorithm performs well on several benchmarks compared to other community detection methods~\cite{lancichinetti2008}.

\section{HCM and HCM*}\label{sec:hcm}
Given a real-world network and the collection of its communities, obtained e.g., using a community detection algorithm, we construct HCM and HCM* in the following way. First, rewire the edges between different communities, using the switching algorithm. Select two inter-community edges uniformly at random, $\{u,v\}$ and $\{w,x\}$. Now delete these edges and replace them by $\{u,x\}$, $\{w,v\}$ if this results in a simple graph. Otherwise keep the original edges $\{u,v\}$ and $\{w,x\}$.  This randomizes the inter-community edges uniformly if this procedure is repeated at least $100E$ times, where $E$ is the number of inter-community edges~\cite{milo2003}. This creates HCM.

To create HCM*, the edges within the communities are also randomized after rewiring the inter-community edges, again using the switching algorithm. This is repeated for all communities.

Now we analyze HCM in more detail, to analytically derive the size of its largest component as in~\cite{stegehuis2015}. 
Let $s_i$ be the size of community $i$, and $k_i$ the number of half-edges from community $i$ to other communities. We call $k_i$ the \textit{inter-community degree} of community $i$.
We define the joint distribution $p_{k,s}$ to be the fraction of communities of size $s$ with inter-community degree $k$. 
We define two distributions and their probability generating functions to calculate the size of the largest component. 
The \textit{excess inter-community degree distribution}
\begin{equation}
q_{k,s}=\frac{(k+1)p_{k+1,s}}{\mean{k}},
\end{equation}
can be interpreted as the probability to arrive in a community with inter-community degree $k$ and size $s$ when traversing a random inter-community edge, excluding the traversed edge. Here $\mean{k}=\sum_{k,s}kp_{k,s}$ is the expected value of $k$.
Similarly, define 
\begin{equation}
r_{k,s}=\frac{sp_{k,s}}{\mean{s}}
\end{equation}
as the probability that a randomly chosen vertex is in a community of size $s$ (including the vertex itself) and has $k$ edges to other communities.
The probability generating functions of these distributions are given by
\begin{align}
g_q(x)&=\sum_{k,s}q_{k,s}x^k=\frac{1}{\mean{k}}\sum_{k,s}kp_{k,s}x^{k-1},\\
g_r(x)&=\sum_{k,s}r_{k,s}x^k=\frac{1}{\mean{s}}\sum_{k,s}sp_{k,s}x^{k}, \label{eq:gr}
\end{align}
and are used to calculate the asymptotic size of the largest component.

Let $u$ be the probability that a community that is reached by traversing a random inter-community edge is not in the giant component, in which case all the communities connected to it cannot be in the giant component either. The $k$ neighboring communities of the reached community are not in the giant component with probability $u^k$. Hence, a community is not in the giant component with probability
\begin{equation}\label{eq:u}
u=\sum_{k,s}q_{k,s}u^k=g_q(u).
\end{equation}
The probability that a randomly chosen vertex is not in the giant component is $\sum_{k,s}r_{k,s}u^k=g_r(u)$. Thus, the proportion of vertices in the largest component $S$ satisfies
\begin{equation}\label{eq:S}
S=1-g_r(u).
\end{equation}
Equations~\eqref{eq:u}\--\eqref{eq:S} can be solved together to find the asymptotic size of the largest component.

Equations~\eqref{eq:u}\--\eqref{eq:S} only depend on the community sizes and the number of edges to other communities. Therefore, when the communities of HCM* are connected, they give the size of the largest component both for HCM and HCM*. In most instances, the number of edges from one vertex of the community to other vertices of the community is large, so HCM* typically generates connected communities, and indeed~\eqref{eq:S} can also be used to calculate the size of the giant component in HCM*.

\subsection{Assortativity of HCM}\label{sec:assort}
The assortativity of the HCM can be computed analytically using~\eqref{eq:assort}. We denote the degree of a randomly chosen vertex among the $N$ vertices of the graph by $D_N$. Then,~\eqref{eq:assort} can be rewritten as
\begin{equation}\label{eq:ass2}
\begin{aligned}[b]
r(G)&=\frac{2\sum_{\{i,j\}\in E}d_id_j-\frac{1}{2L}\left(\sum_i d_i^2\right)^2}{\sum_id_i^3-\frac{1}{2L}\left(\sum_i d_i^2\right)^2}
=\frac{\frac{2}{N}\sum_{\{i,j\}\in E}d_id_j-\frac{\left(\frac{1}{N}\sum_i d_i^2\right)^2}{\frac{1}{N}\sum_i d_i}}{\frac{1}{N}\sum_id_i^3-\frac{\left(\frac{1}{N}\sum_i d_i^2\right)^2}{\frac{1}{N}\sum_i d_i}}\\
&=\frac{\frac{2}{N}\sum_{\{i,j\}\in E}d_id_j-\frac{\mathbb{E}[D_N^2]^2}{\mathbb{E}[D_N]}}{\mathbb{E}[D_N^3]-\frac{\mathbb{E}[D_N^2]^2}{\mathbb{E}[D_N]}}.
\end{aligned}
\end{equation}
Therefore, the only term of assortativity that depends on the community structure of HCM is the first term in the numerator. 
The edges of HCM can be split into two sets: the edges that are entirely inside a community, and the edges that are between two different communities, denoted by $E_c$ and $E_b$ respectively. The edges inside communities are fixed given the community shape. Let the $n$ communities of the network be denoted by $\{H_1,\dots,H_n\}$. For a given community $H$, let $Q(H)$ denote
\begin{equation}
Q(H)=\sum_{\{i,j\}\in E_H}d_id_j.
\end{equation}
Then the contribution of the intra-community edges to the first term in the numerator can be written as

\begin{equation}\label{eq:ascom}
\mathbb{E}\Big[\frac{1}{N}\sum_{\{i,j\}\in E_c}d_id_j\Big]=\frac{1}{N}\sum_{k=1}^n\sum_{\{i,j\}\in E_{H_k}}d_id_j=\frac{1}{n\mathbb{E}[S_n]}\sum_{k=1}^nQ(H_k)=\frac{\mathbb{E}[Q_n]}{\mathbb{E}[S_n]},
\end{equation}
where $\mathbb{E}[Q_n]$ is the expected value of $Q$ of a randomly chosen community, and $\mathbb{E}[S_n]$ the size of a uniformly chosen community.
Let $D_N^{\sss{(b)}}$ denote the number of edges to other communities of a randomly chosen vertex, and $L^{\sss{(b)}}$ the total number of edges between communities. The probability that a specific half-edge will be paired with another specific half-edge equals $1/(2L^{\sss{(b)}}-1)$, since the half-edges are paired at random. We denote the number of half-edges adjacent to vertex $i$ by $d_i^{\sss{(b)}}$. Then the contribution of the inter-community edges can be written as
\begin{equation}\label{eq:asconf}
\begin{aligned}[b]
\mathbb{E}\Big[\frac{1}{N}\sum_{\{i,j\}\in E_b}d_id_j\Big]&=\frac{1}{N}\sum_{i=1}^N\sum_{j=1}^N\sum_{k=1}^{d_i^{\sss{(b)}}}\sum_{l=1}^{d_j^{\sss{(b)}}}\frac{d_id_j}{2L^{\sss{(b)}}-1}=
\frac{1}{N}\sum_{i=1}^N\sum_{j=1}^N\frac{d_id_i^{\sss{(b)}}d_jd_j^{\sss{(b)}}}{2L^{\sss{(b)}}-1}\\
&\approx\frac{1}{N2L^{\sss{(b)}}}\left(\sum_{i=1}^Nd_id_i^{\sss{(b)}}\right)^2=\frac{1}{2L^{\sss{(b)}}/N}\left(\frac{1}{N}\sum_{i=1}^Nd_id_i^{\sss{(b)}}\right)^2=\frac{\mathbb{E}[D_ND_N^{\sss{(b)}}]^2}{\mathbb{E}[D_N^{\sss{(b)}}]}.
\end{aligned}
\end{equation}
Combining~\eqref{eq:ass2},~\eqref{eq:ascom} and~\eqref{eq:asconf} gives for the expected assortativity of a HCM network that
\begin{equation}
\mathbb{E}[r(G)]=\frac{2\frac{\mathbb{E}[D_ND_N^{\sss{(b)}}]^2}{\mathbb{E}[D_N^{\sss{(b)}}]}+2\frac{\mathbb{E}[Q_n]}{\mathbb{E}[S_n]}-\frac{\mathbb{E}[D_N^2]^2}{\mathbb{E}[D_N]}}{\mathbb{E}[D_N^3]-\frac{\mathbb{E}[D_N^2]^2}{\mathbb{E}[D_N]}}.
\end{equation}

Supplementary Table~\ref{tab:assort} shows that HCM and HCM* generate networks that match the assortativity of the original network closer than a configuration model. 
However, the assortativity generated by HCM does not always match its theoretical value. An explanation for this is that HCM generates simple graphs, while the theoretical estimate does not take this into account. Since both ends of a self-loop have the same degree, having non-simple graphs increases the assortativity. Furthermore, these self-loops typically occur at nodes of large degree, increasing the assortativity even further, so that the theoretical assortativity is higher than the observed assortativity. 
%However, we see that HCM and HCM* in most cases generate networks with higher assortativity than the original network. This is because the original networks are simple graphs, whereas HCM and HCM* do not need to be simple. Therefore, HCM and HCM* both contain self-loops. Since both ends of the self-loop have the same degree, this increases the assortativity. Furthermore, these self-loops typically occur at nodes of large degree, increasing the assortativity even further. 

\subsection{Overlap of HCM and original communities}
By keeping the sets of communities fixed, we expect both HCM and HCM* to generate networks with a similar community structure as the original data set. To test how similar the community structures of the generated networks and the original networks are, we define the similarity $w_{\text{HCM}}$ of the two community structures as
\begin{equation}\label{eq:sim}
w_{\text{HCM}}=\frac{1}{N}\sum_i\frac{|\mathcal{C}_i\cap\mathcal{C}_i^{\sss{\text{HCM}}}|}{|\mathcal{C}_i|},
\end{equation}
where $\mathcal{C}_i$ and $\mathcal{C}_i^{\sss{\text{HCM}}}$ are the sets of vertices that are in the same community as vertex $i$ in the original network, and the HCM network respectively. We define the similarity for the community structure generated by HCM* similarly as $w_{\text{HCM*}}$.
Table~\ref{tab:con} presents this similarity measure for all networks. We see that for most networks, the degree of overlap is large, but for the AS network, the overlap between the original network community sets and the networks generated by HCM or HCM* is smaller. This may be explained by the fact that the AS network has less dense communities, so that rewiring the edges between communities can easily shift vertices from one community to the other.

\subsection{Connectedness of HCM* communities}
The communities of HCM* are generated by rewiring the edges of the original communities. This may cause the communities to be disconnected after rewiring. Table~\ref{tab:con} presents the fraction of disconnected communities $f_{\text{dis}}$ that HCM* generates. Table~\ref{tab:con} also presents $N_{\text{dis}}$, the average number of vertices that are not connected to the largest component of the community after rewiring, given that the community is disconnected. We see that the fraction of disconnected communities  is different for the different networks. For the networks with a more dense community structure, the probability that a community becomes disconnected after rewiring is low, while for for example the AS network this probability is higher. In all cases, the number of vertices that are disconnected from the largest component is low, indicating that the community stays largely connected.

\section{Types of epidemic processes}\label{sec:perc}

\begin{figure}[!h]
	\centering
	\subfloat[]{
		\centering
		\includegraphics[ width=0.3\textwidth]{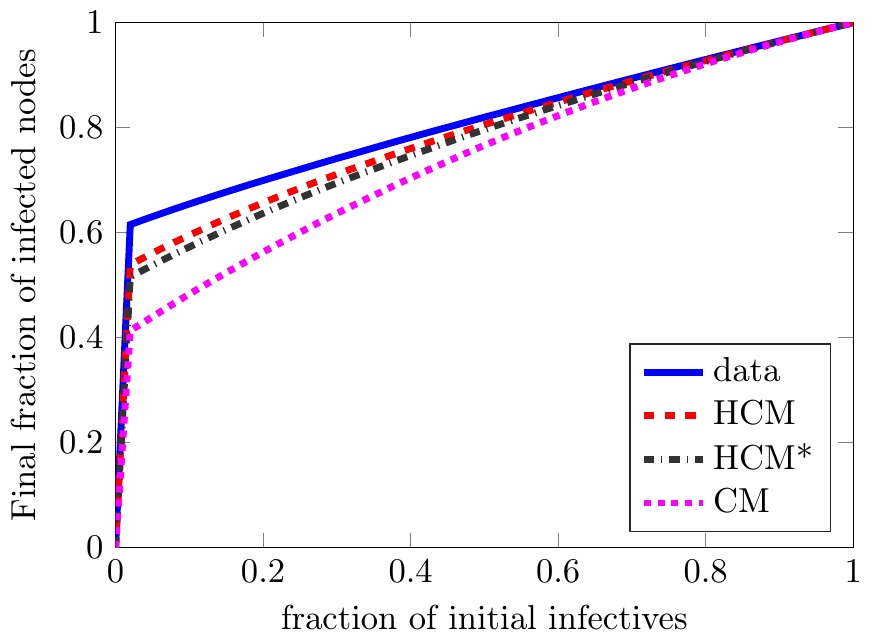}
		
		\label{fig:bootas}
	}
	\subfloat[]{
		\centering
		\includegraphics[ width=0.3\textwidth]{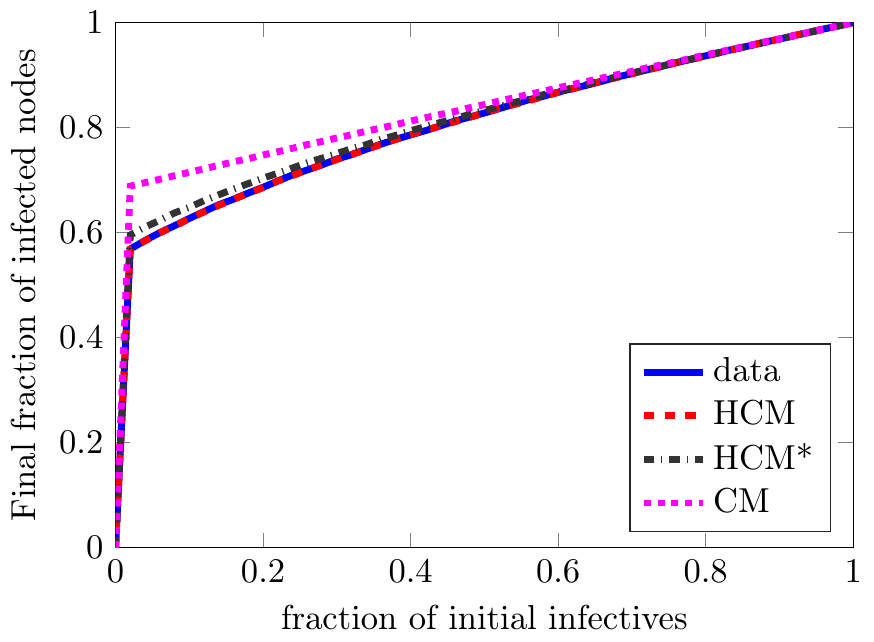}
		
		\label{fig:booten}
	}
	\subfloat[]{
		\centering
		\includegraphics[ width=0.3\textwidth]{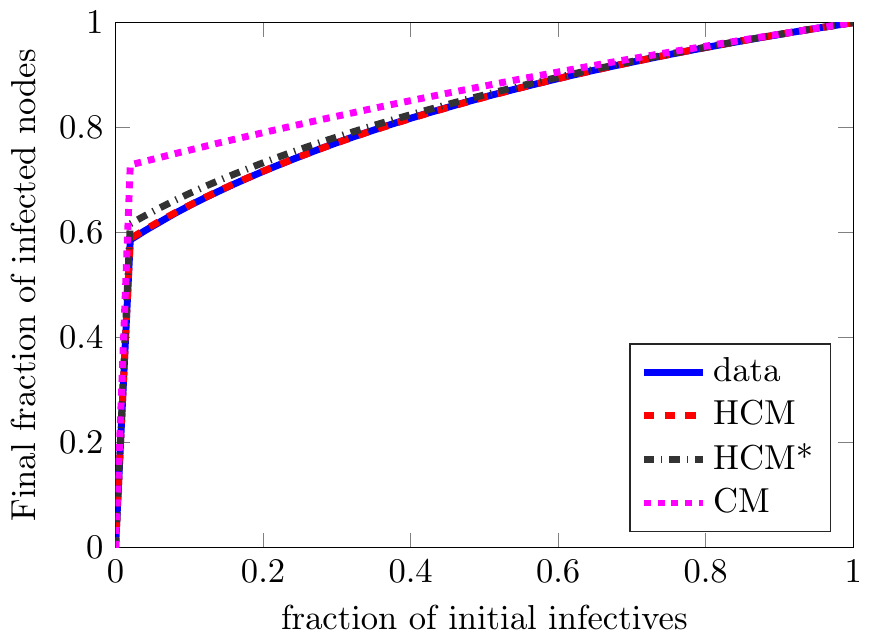}
		
		\label{fig:boothep}
	}

	\subfloat[]{
		\centering
		\includegraphics[ width=0.3\textwidth]{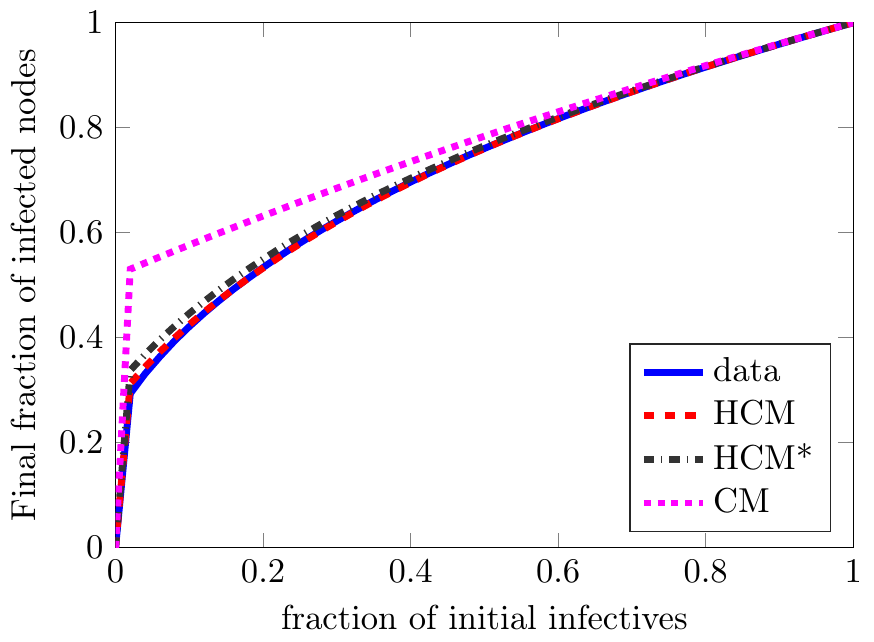}
		
		\label{fig:bootPGP}
	}
	\subfloat[]{
		\centering
		\includegraphics[ width=0.3\textwidth]{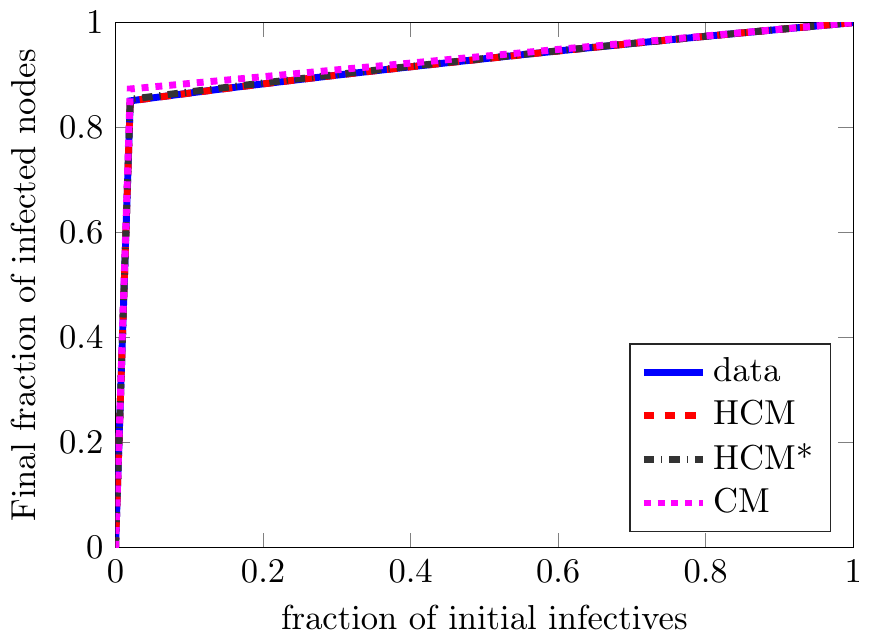}
		
		\label{fig:bootpow}
	}
	\subfloat[]{
		\centering
		\includegraphics[ width=0.3\textwidth]{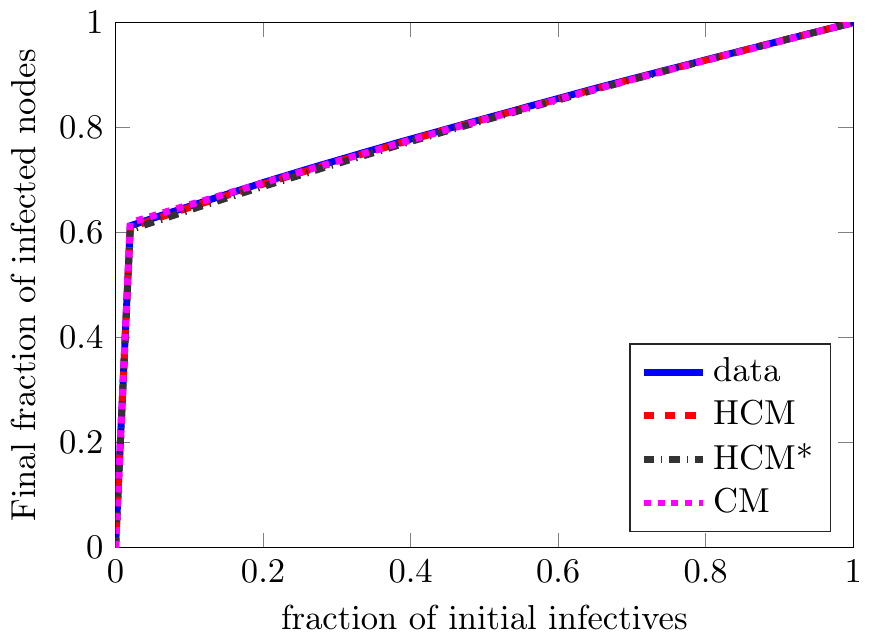}
		
		\label{fig:bootyeast}
	}
	\caption{\textbf{HCM, HCM* and CM under bootstrap percolation compared to real-world networks.} a) Autonomous Systems network b) \textsc{Enron} email network c) Collaboration network in High energy physics d) PGP network e) \textsc{Facebook} friendship network f) yeast network. Initially, a certain fraction of the vertices is infected at random. Then, a vertex becomes infected when at least 2 of its neighbors are infected. The final fraction of infected vertices is the average of 500 generated graphs.}
	\label{fig:boot}	
\end{figure}

\begin{figure}[!h]
	\centering
	\subfloat[]{
		\centering
		\includegraphics[ width=0.3\textwidth]{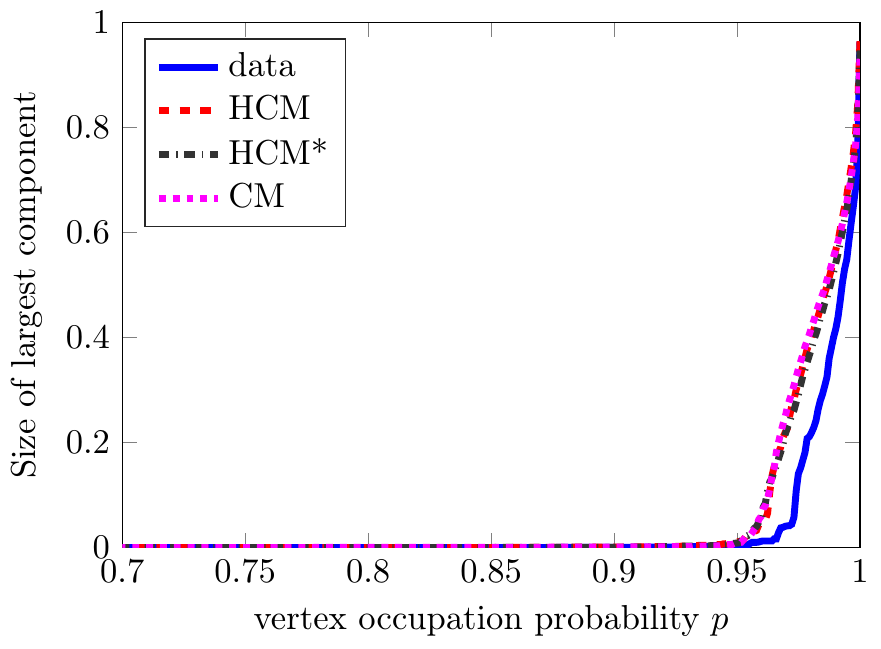}
		
		\label{fig:percdas}
	}
	\subfloat[]{
		\centering
		\includegraphics[ width=0.3\textwidth]{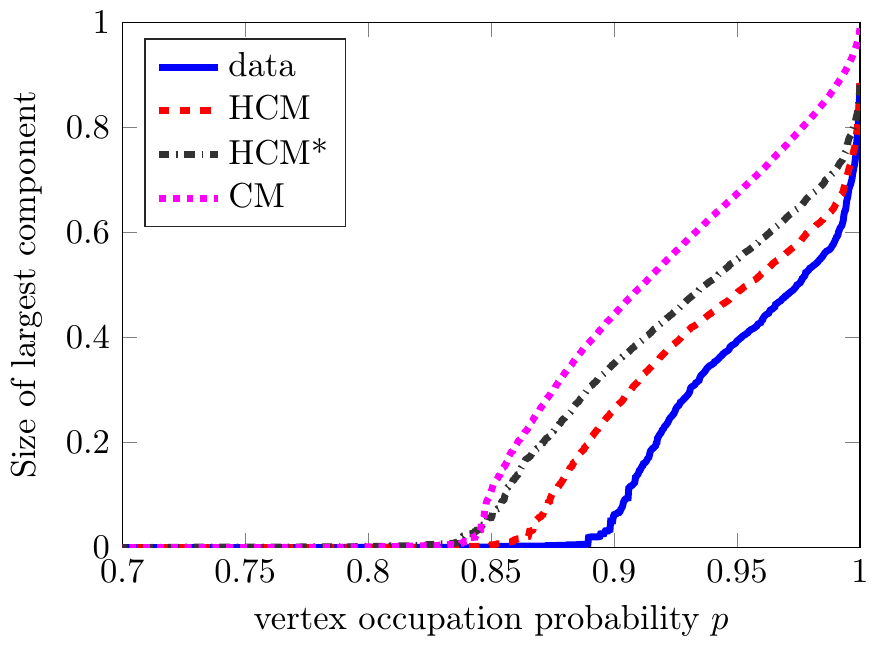}
		\label{fig:percden}
	}
	\subfloat[]{
		\centering
		\includegraphics[ width=0.3\textwidth]{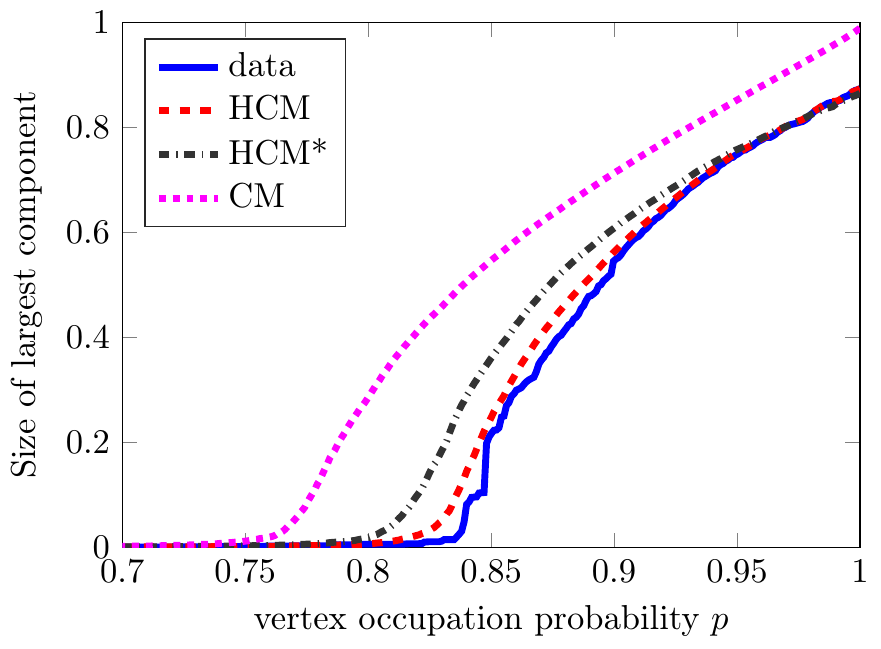}	 
		\label{fig:percdHEP}
	}

	\subfloat[]{
		\centering
		\includegraphics[ width=0.3\textwidth]{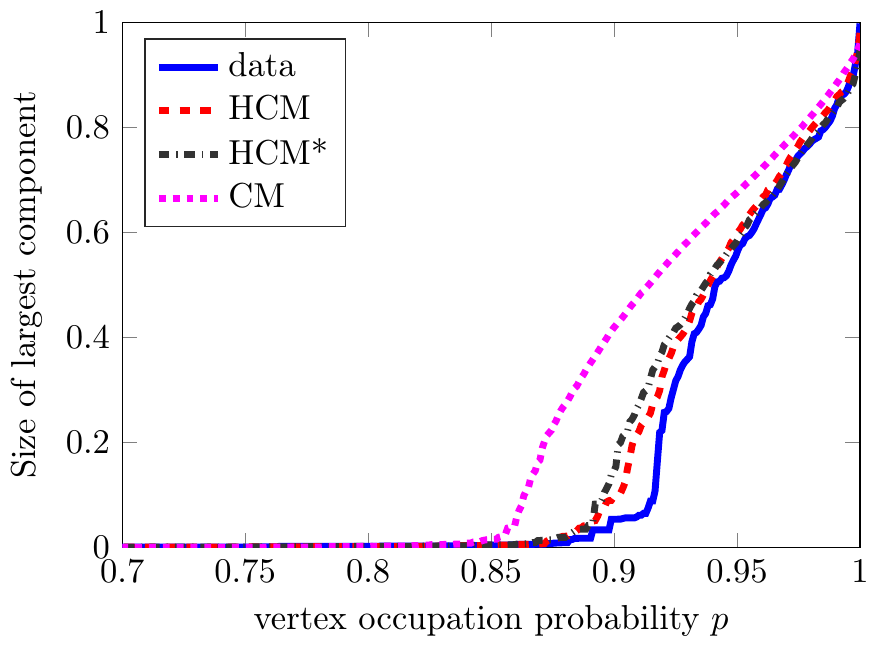}
		
		\label{fig:percdPGP}
	}
	\subfloat[]{
		\centering
		\includegraphics[ width=0.3\textwidth]{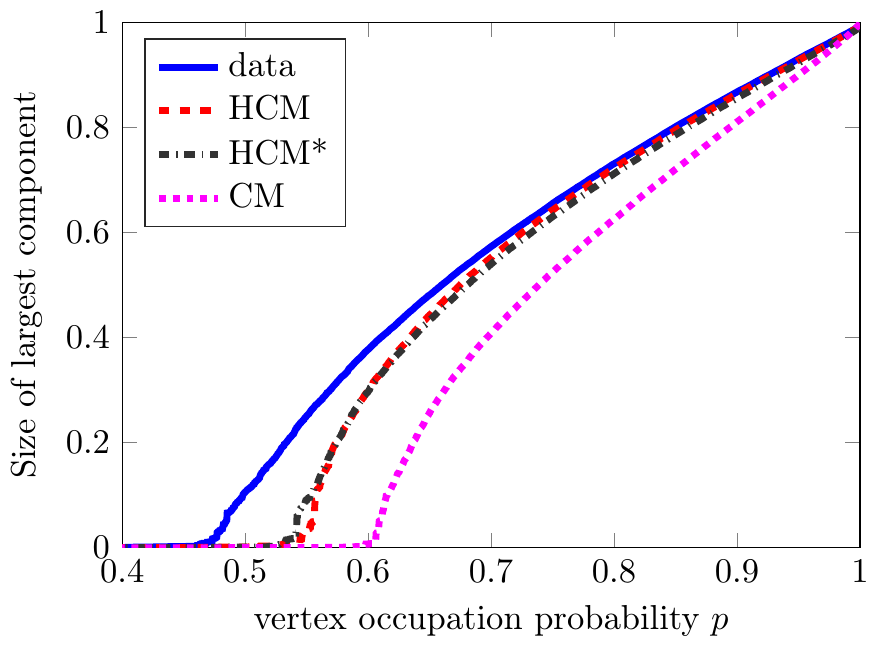}
		
		\label{fig:percdpow}
	}
	\subfloat[]{
		\centering
		\includegraphics[ width=0.3\textwidth]{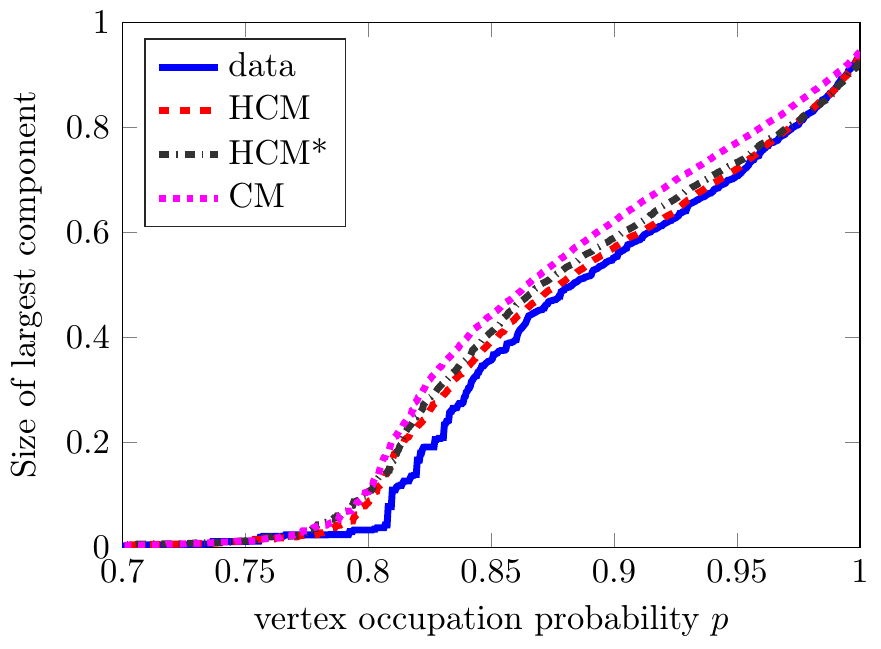}
		
		\label{fig:percdyeast}
	}
	\caption{\textbf{HCM, HCM* and CM under a targeted attack, compared to real-world networks.} a) Autonomous Systems network b) \textsc{Enron} email network c) Collaboration network in High energy physics d) PGP network e) \textsc{Facebook} friendship network f) yeast network. The fraction of $1-p$ vertices of highest degree are removed. The size of the largest component after the vertices are removed is the average of 500 generated graphs.}
	\label{fig:percd}	
\end{figure}

\begin{figure}[!h]
	\centering
	\subfloat[]{
		\centering
		\includegraphics[ width=0.3\textwidth]{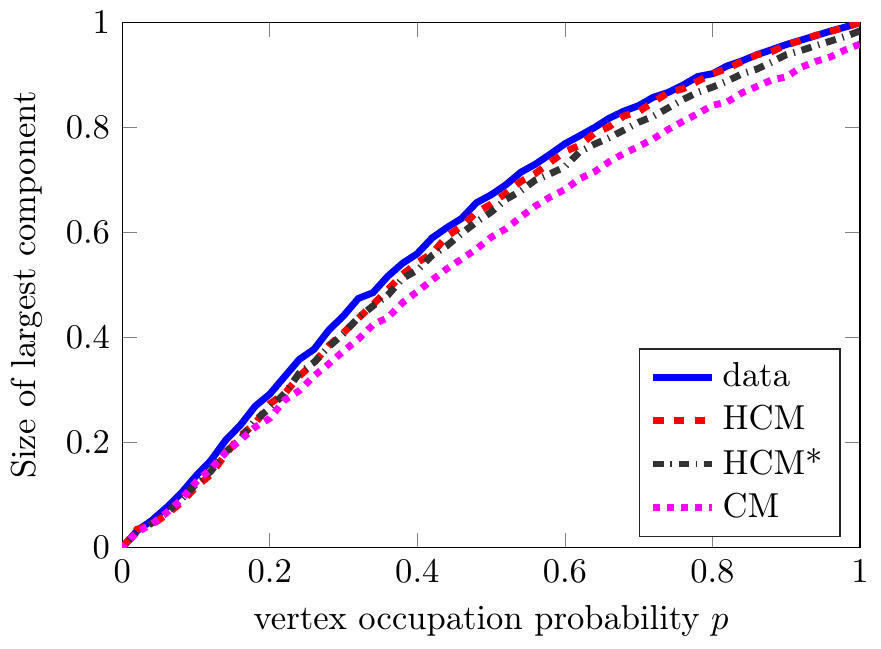}
		
		\label{fig:percsas}
	}
	\subfloat[]{
		\centering
		\includegraphics[ width=0.3\textwidth]{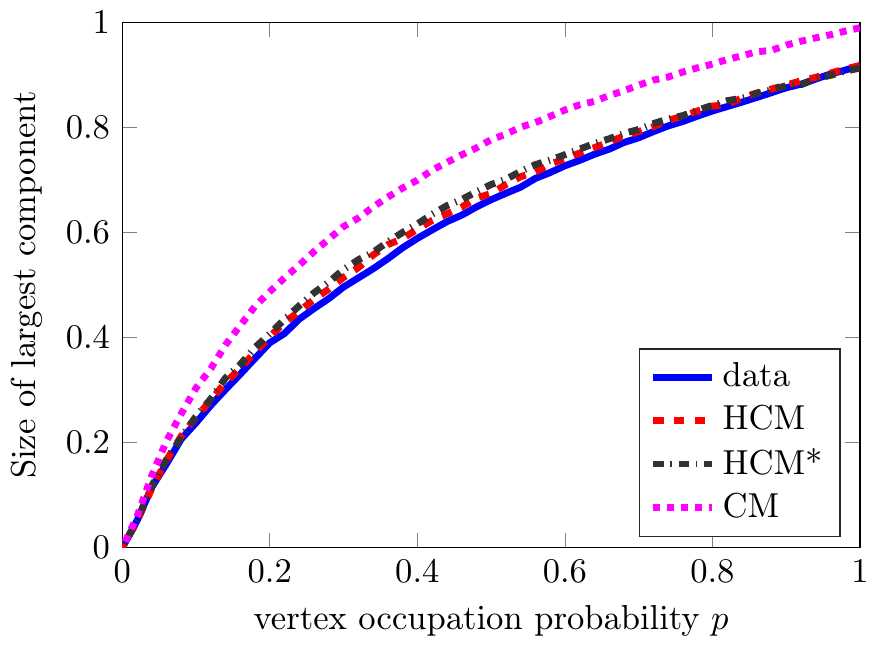}
		
		\label{fig:percsen}
	}
	\subfloat[]{
		\centering
		\includegraphics[ width=0.3\textwidth]{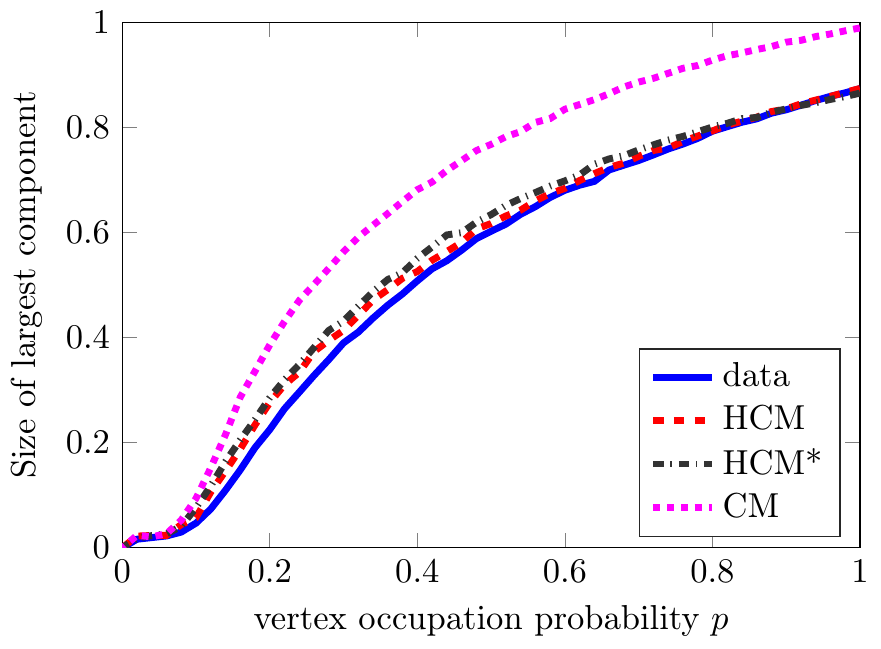}
		\label{fig:percsHEP}
	}

	\subfloat[]{
		\centering
		\includegraphics[ width=0.3\textwidth]{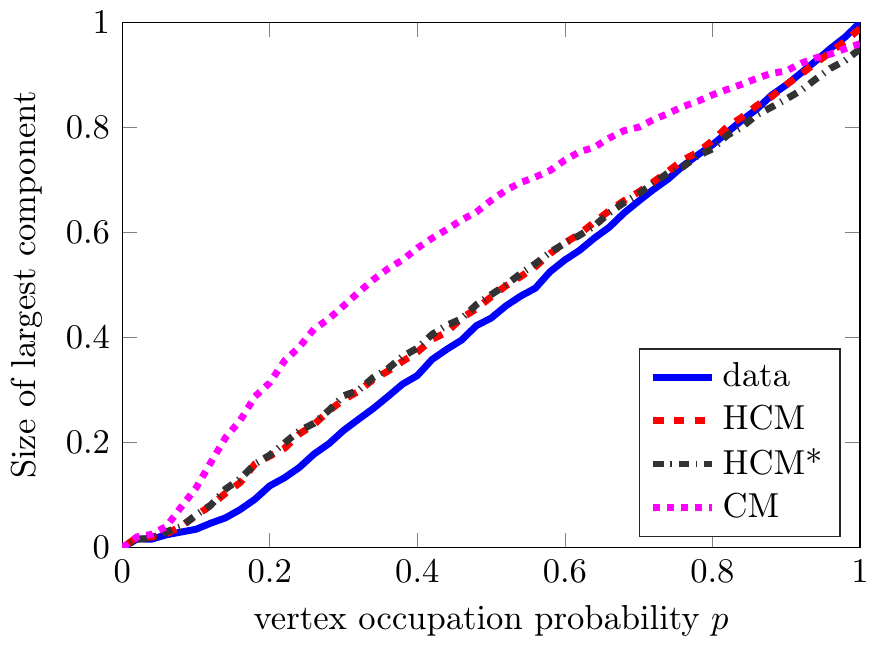}
		
		\label{fig:percsPGP}
	}
	\subfloat[]{
		\centering
		\includegraphics[ width=0.3\textwidth]{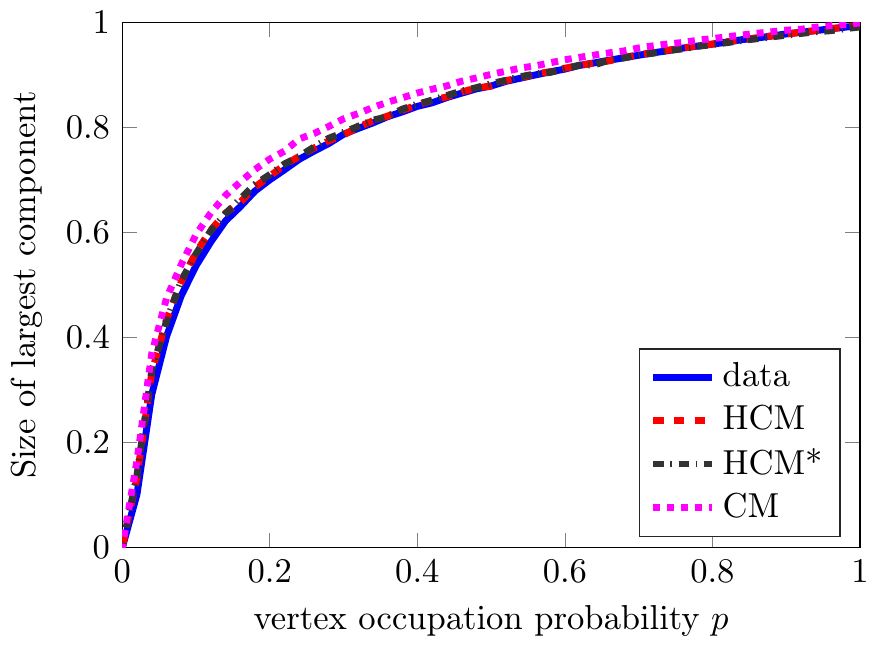}
		
		\label{fig:percspow}
	}
	\subfloat[]{
		\centering
		\includegraphics[ width=0.3\textwidth]{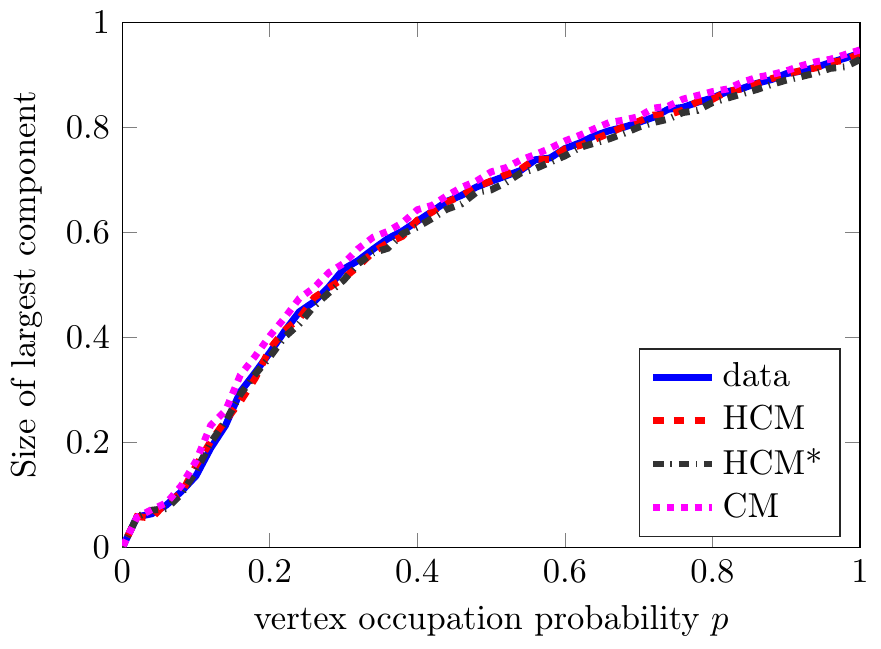}
		
		\label{fig:percsyeast}
	}
	\caption{\textbf{HCM, HCM* and CM under site percolation compared to real-world networks.} a) Autonomous Systems network b) \textsc{Enron} email network c) Collaboration network in High energy physics d) PGP network e) \textsc{Facebook} friendship network f) yeast network. Independently, every vertex is removed from the network with probability $1-p$. The size of the largest component after the vertices are removed is the average of 500 generated graphs.}
	\label{fig:percs}	
\end{figure}

\begin{figure}[!h]
	\centering
	\subfloat[]{
		\centering
		\includegraphics[ width=0.3\textwidth]{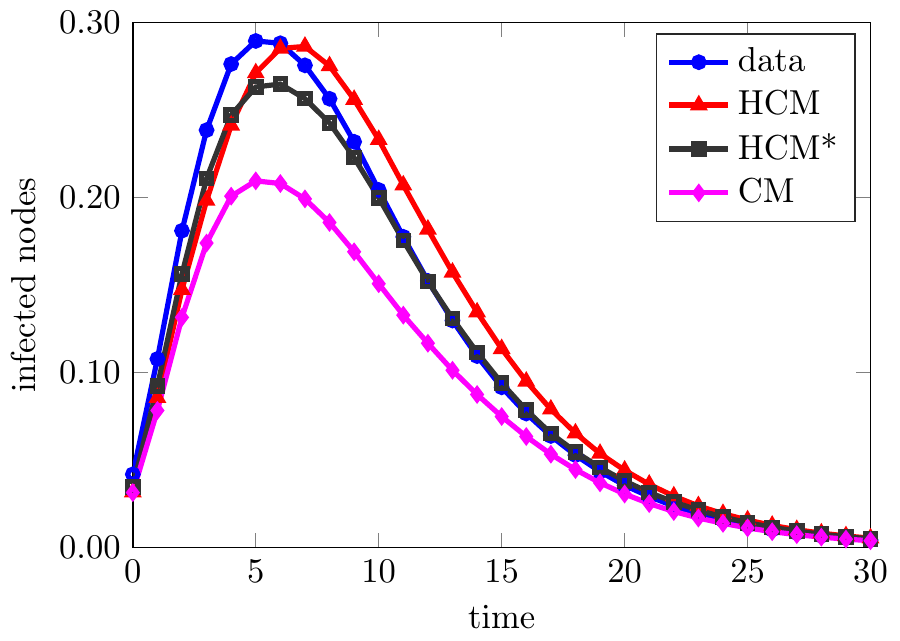}
		
		\label{fig:infas}
	}
	\subfloat[]{
		\centering
		\includegraphics[ width=0.3\textwidth]{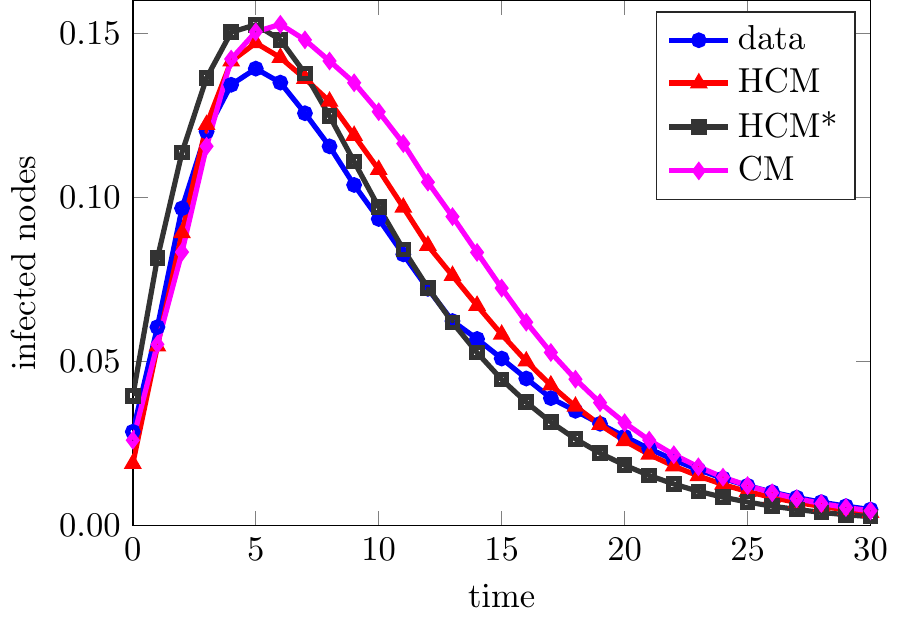}
		
		\label{fig:infen}
	}
	\subfloat[]{
		\centering
		\includegraphics[ width=0.3\textwidth]{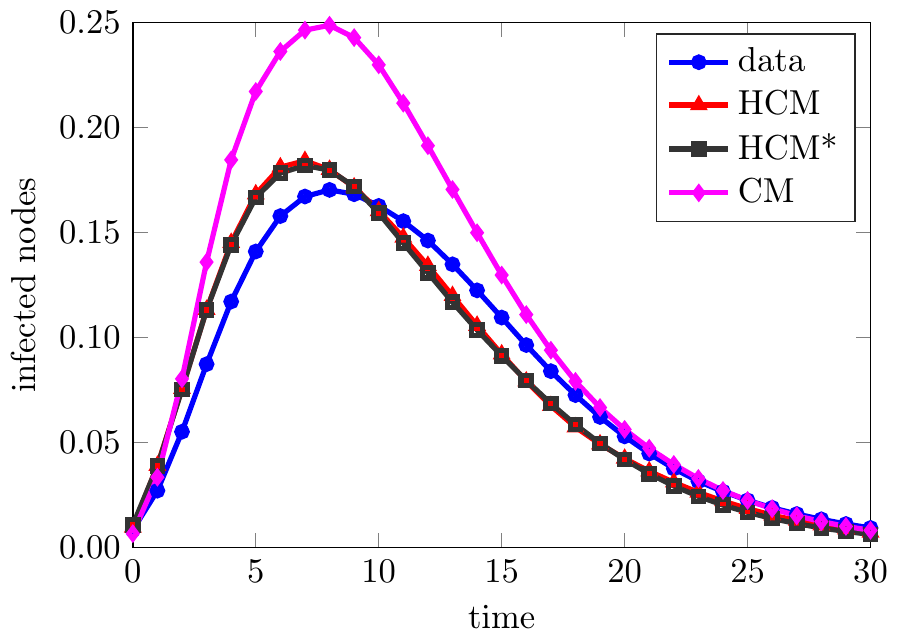}
		\label{fig:infHEP}
	}

	\subfloat[]{
		\centering
		\includegraphics[ width=0.3\textwidth]{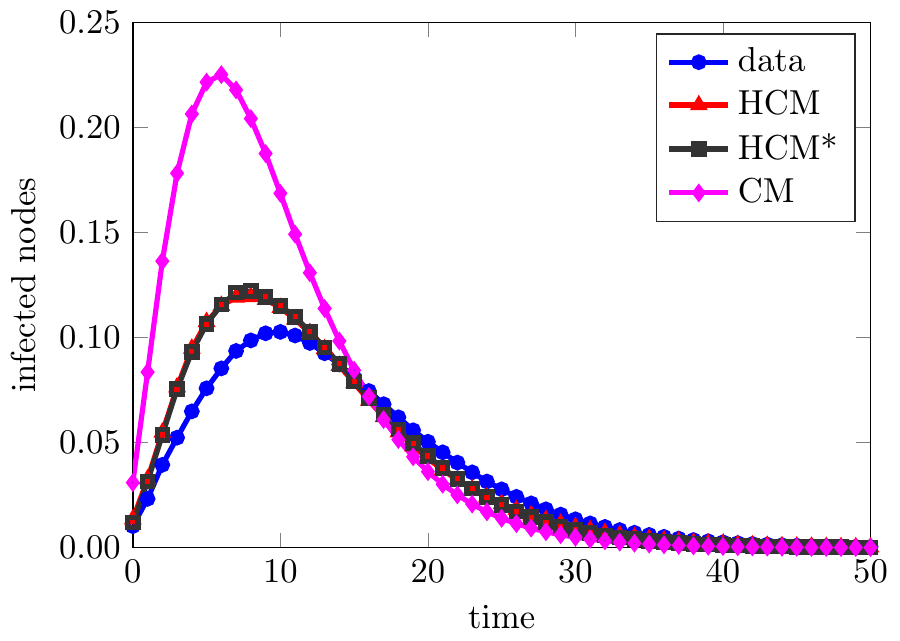}
		
		\label{fig:infPGP}
	}
	\subfloat[]{
		\centering
		\includegraphics[ width=0.3\textwidth]{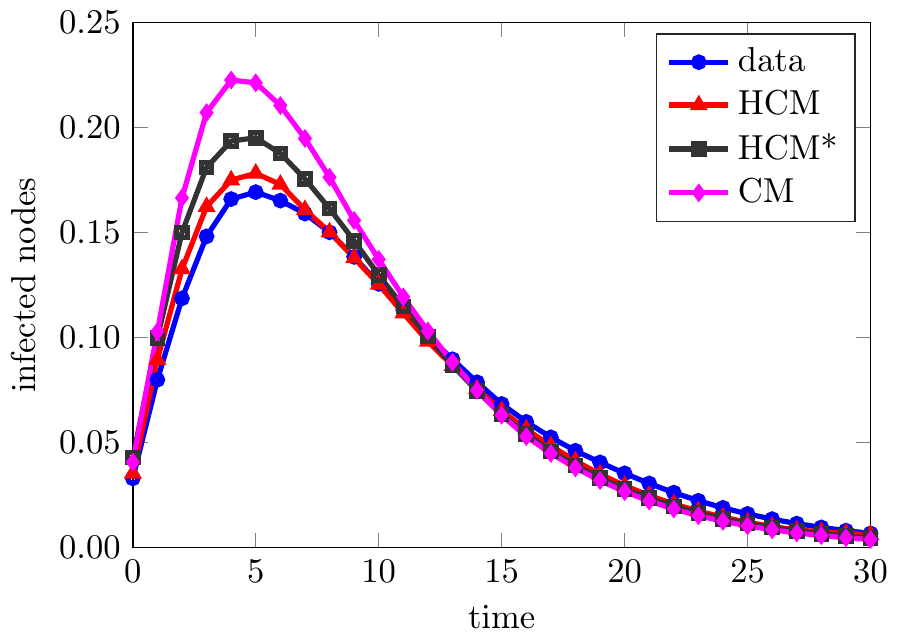}
		
		\label{fig:infpow}
	}
	\subfloat[]{
		\centering
		\includegraphics[ width=0.3\textwidth]{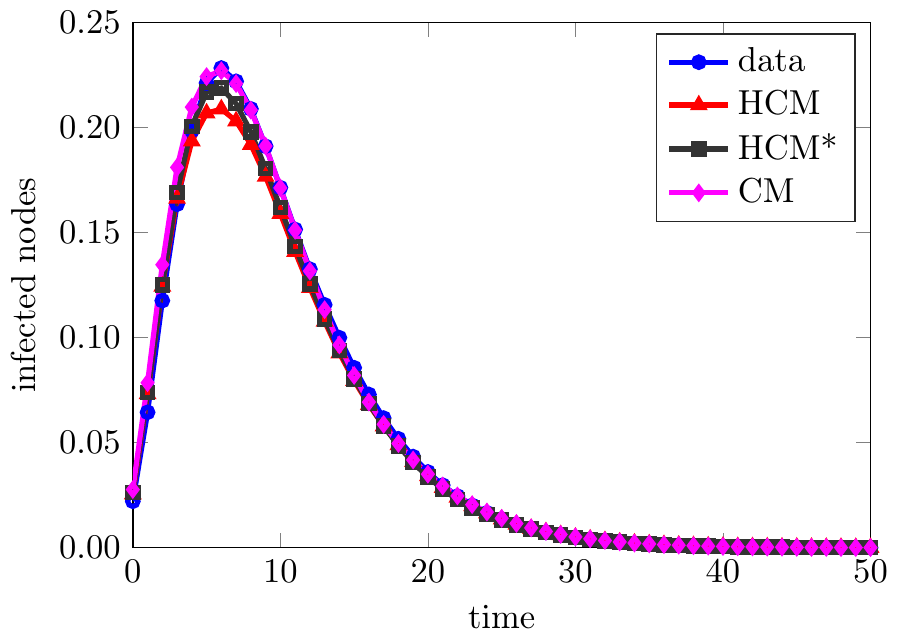}
		
		\label{fig:infyeast}
	}
	\caption{\textbf{The number of infected individuals in an SIR epidemic in HCM, HCM* and CM compared to real-world networks.}  a) Autonomous Systems network b) \textsc{Enron} email network c) Collaboration network in High energy physics d) PGP network e) \textsc{Facebook} friendship network f) yeast network. The presented results are the average of 500 generated graphs, with recovery rate $\gamma=1$ and infection rate $\beta=3\mean{d}/\gamma$, where $\mean{d}$ is the mean degree.}
	\label{fig:inf}	
\end{figure}

\begin{figure}[!h]
	\centering
	\subfloat[]{
		\centering
		\includegraphics[ width=0.3\textwidth]{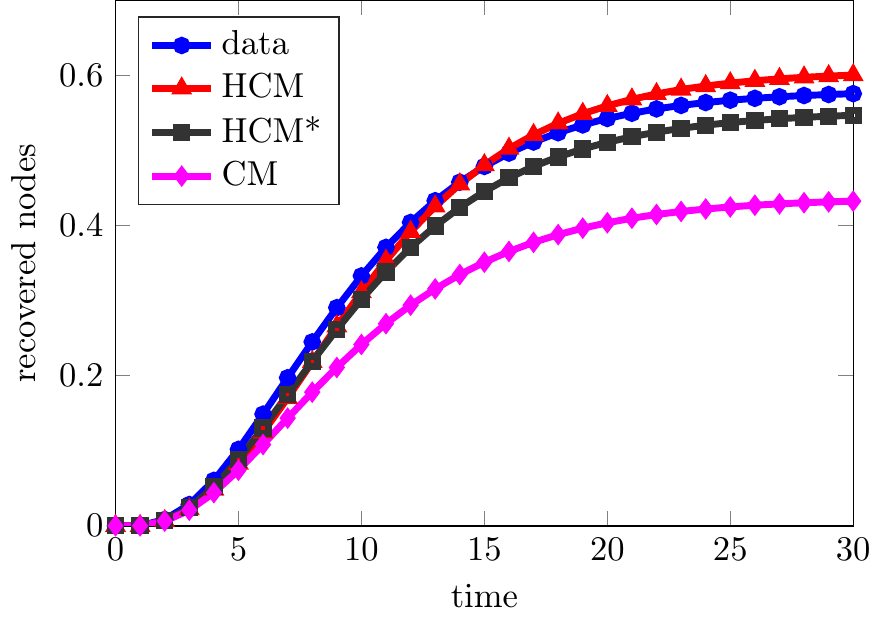}
		
		\label{fig:recas}
	}
	\subfloat[]{
		\centering
		\includegraphics[ width=0.3\textwidth]{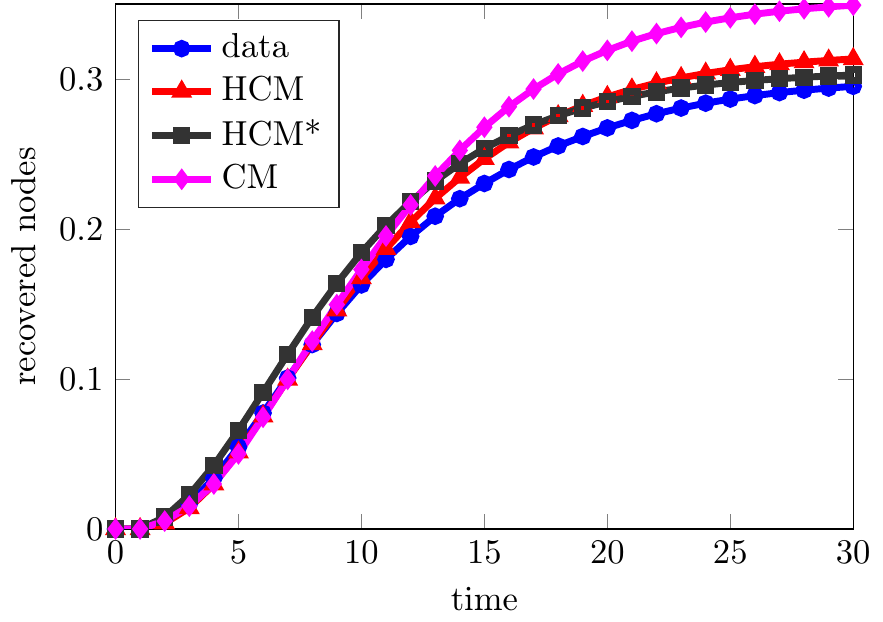}
		
		\label{fig:recen}
	}
	\subfloat[]{
		\centering
		\includegraphics[ width=0.3\textwidth]{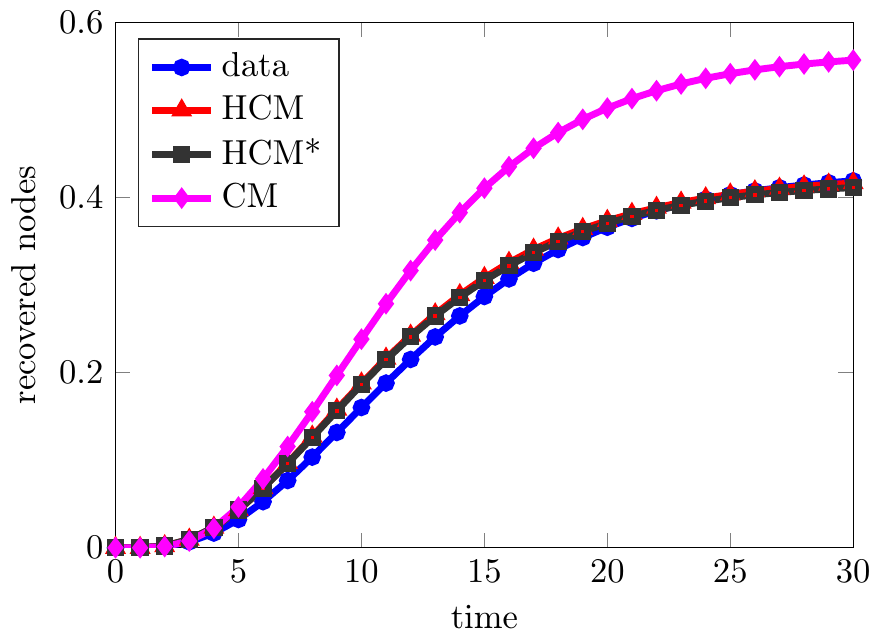}
		\label{fig:recHEP}
	}

	\subfloat[]{
		\centering
		\includegraphics[ width=0.3\textwidth]{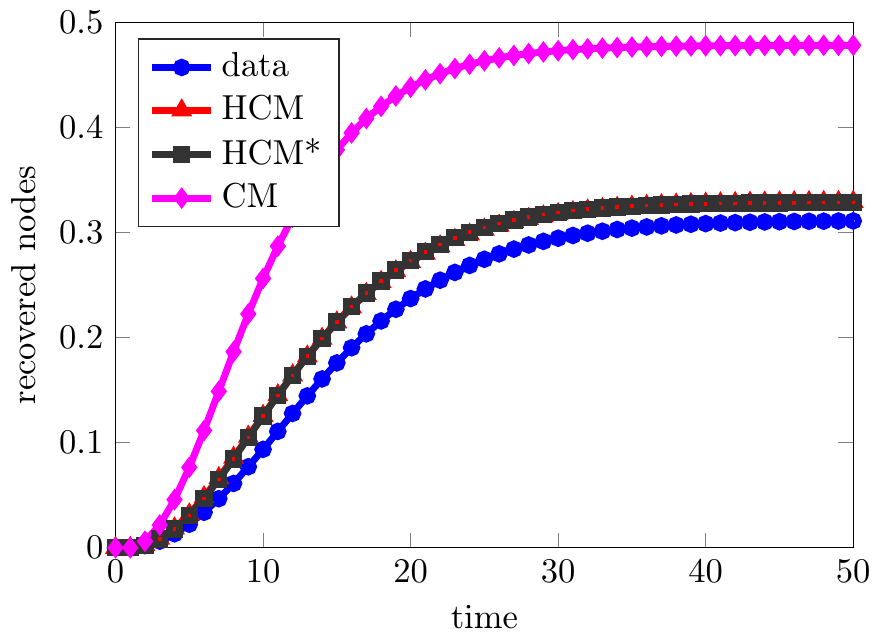}
		
		\label{fig:recPGP}
	}
	\subfloat[]{
		\centering
		\includegraphics[ width=0.3\textwidth]{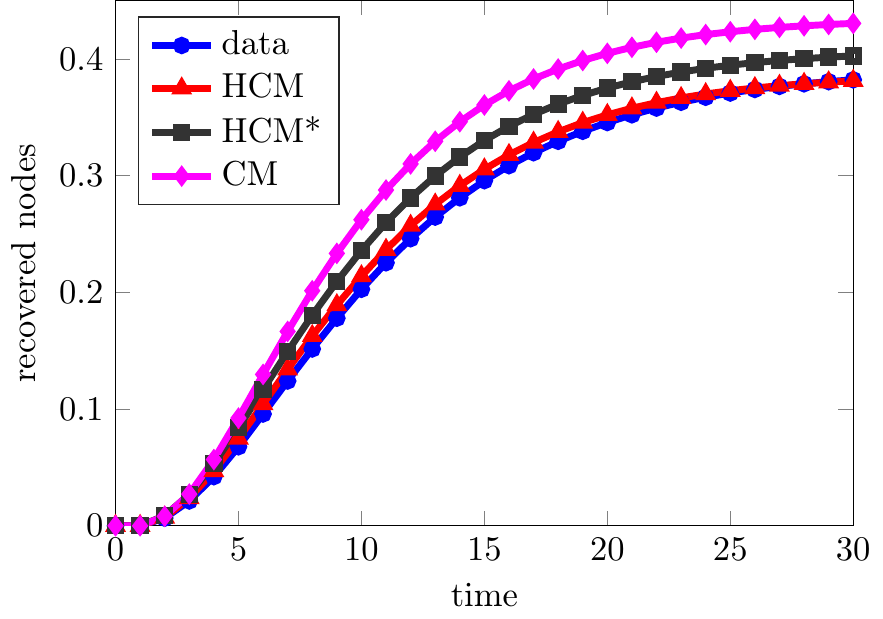}
		
		\label{fig:recpow}
	}
	\subfloat[]{
		\centering
		\includegraphics[ width=0.3\textwidth]{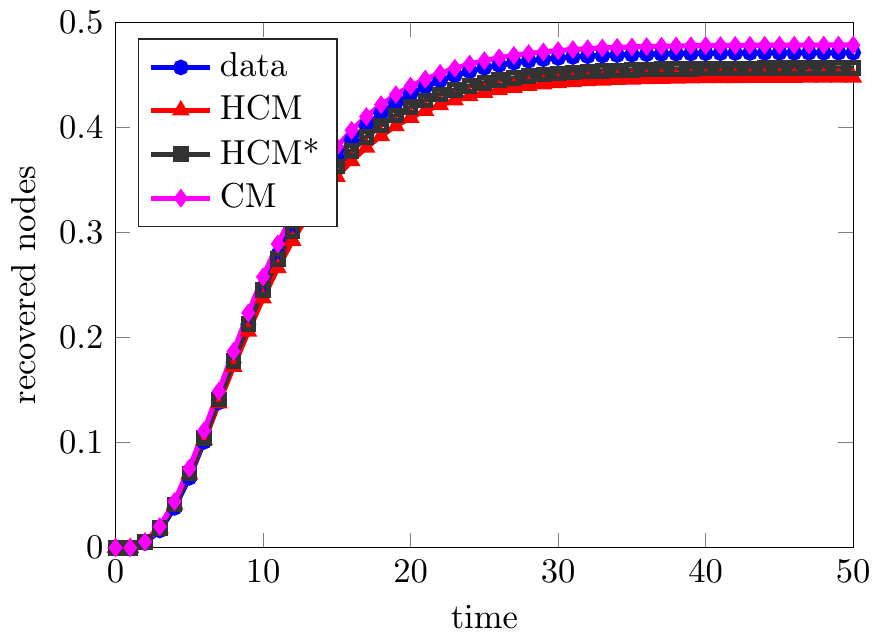}
		
		\label{fig:recyeast}
	}
	\caption{\textbf{The number of recovered individuals in an SIR epidemic in HCM, HCM* and CM compared to real-world networks.} a) Autonomous Systems network b) \textsc{Enron} email network c) Collaboration network in High energy physics d) PGP network e) \textsc{Facebook} friendship network f) yeast network. The presented results are the average of 500 generated graphs, with the recovery rate $\gamma=1$ and the infection rate $\beta=3\mean{d}/\gamma$, where $\mean{d}$ is the mean degree.}
	\label{fig:rec}	
\end{figure}

In Figure~\ref{fig:perc} and Supplementary Figures~\ref{fig:boot}-~\ref{fig:rec}, the results of several epidemic processes are plotted. Here we describe these processes in more detail.
\subsection{Bond percolation}
In bond percolation, every edge of the network is deleted independently with probability $1-p$. The quantity of interest is the fraction of vertices that are in the largest component after this deletion process.  
\subsection{Site percolation}
In site percolation, every vertex, and all edges adjacent to it, are deleted with probability $1-p$, independently for every vertex. As in bond percolation, we are interested in the fraction of vertices in the largest component after this deletion process.
\subsection{Targeted attack}
In a targeted attack, a fraction of $p$ of the vertices and the edges adjacent to them are removed, starting with the highest degree vertex, then the second highest degree vertex and so on. Again, the quantity of interest is the fraction of vertices in the giant component after deleting the edges.
\subsection{Bootstrap percolation}
In bootstrap percolation with threshold $t$ initially a certain fraction of vertices is infective. The initially infected vertices are selected at random. Then, every vertex with at least $t$ infected neighbors also becomes infected. This process continues until no new vertices become infected anymore. In the results, we consider bootstrap percolation with threshold $t=2$. The quantity of interest is the fraction of infected vertices when the process has stopped.

\subsection{SIR epidemic}
In an SIR epidemic, vertices are either susceptible, infected or recovered. One vertex is selected uniformly at random to be the initial infective. Then, every infected vertex infects his susceptible neighbors independently at rate $\beta$. Every infected vertex recovers at rate $\gamma$. As in~\cite{salathe2010}, we set $\gamma=1$ and $\beta=3\mean{d}/\gamma$, where $\mean{d}$ is the average degree of the network. We are interested in how the fraction of infected and recovered vertices evolves over time. Note that since every vertex is either susceptible, infected or recovered, the fraction of susceptible vertices is then also known.

\section*{Supplementary Discussion}\label{sec:data}
\begin{figure}[!h]
	\centering
	\subfloat[]{
		\centering
		\includegraphics[ width=0.3\textwidth]{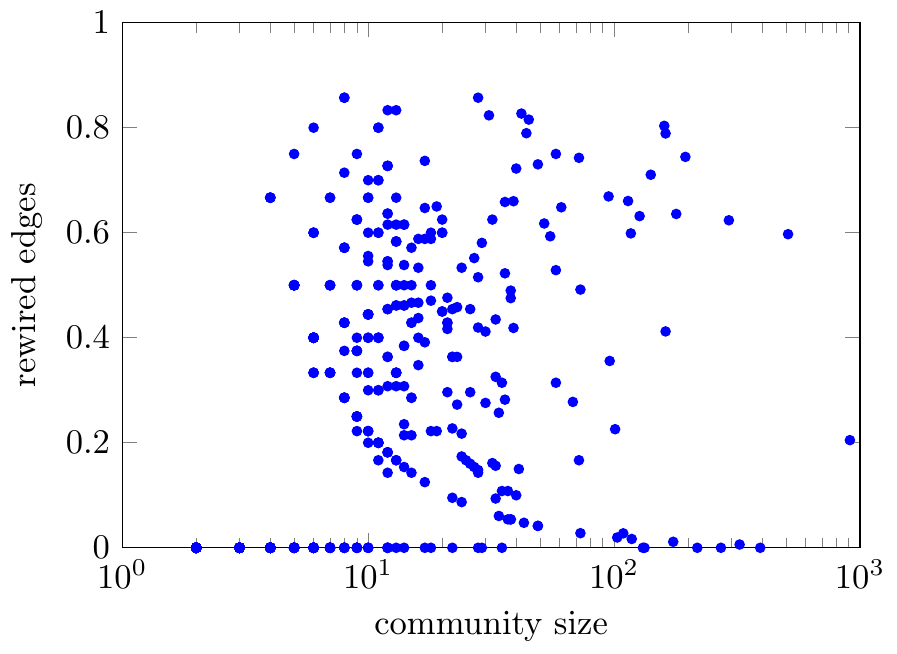}	 
		\label{fig:rewas}
	}
	\subfloat[]{
		\centering
		\includegraphics[ width=0.3\textwidth]{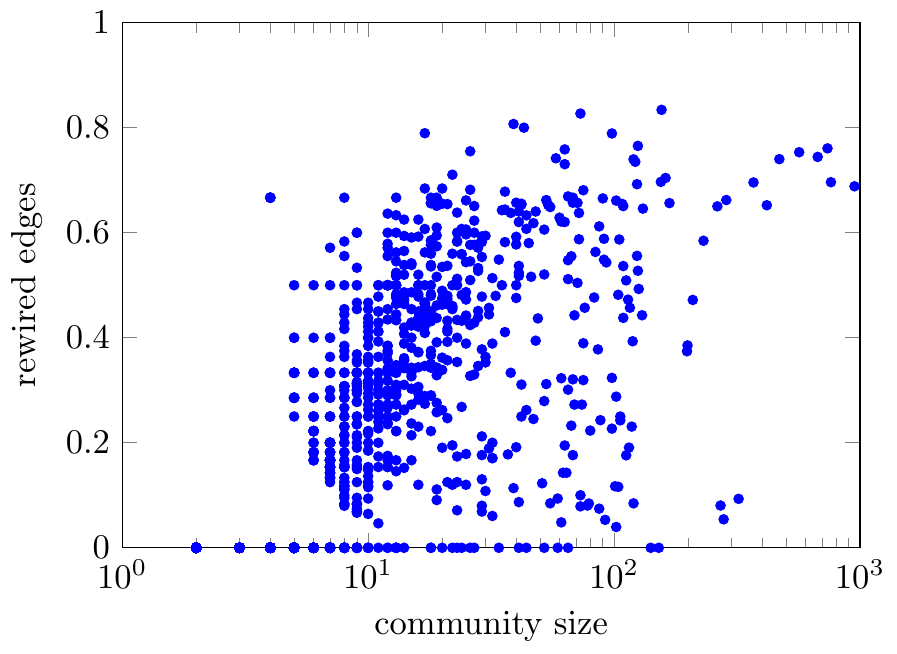}
		
		\label{fig:rewen}
	}
	\subfloat[]{
		\centering
		\includegraphics[ width=0.3\textwidth]{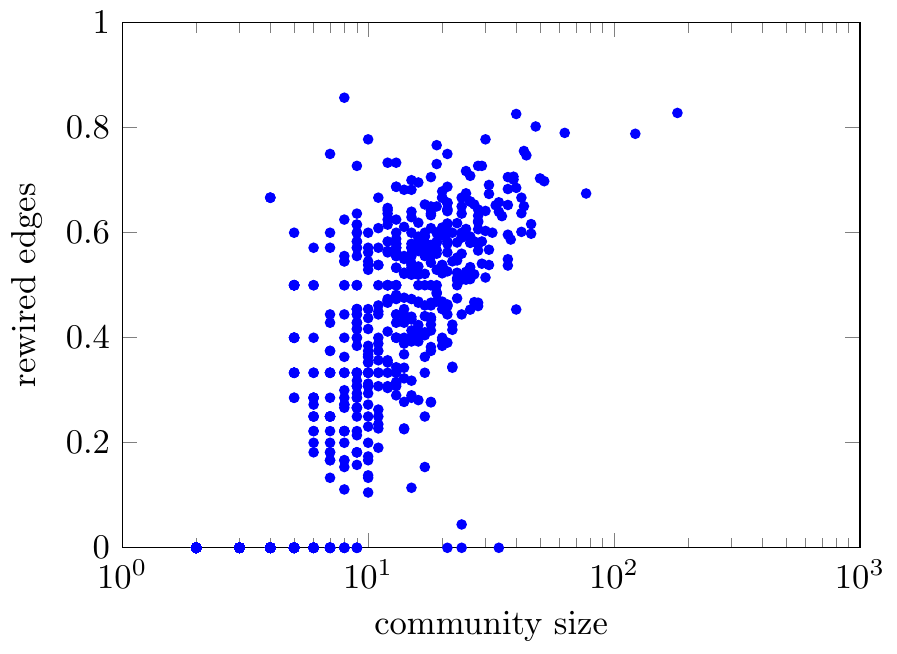}
		
		\label{fig:rewHEP}
	}

	\subfloat[]{
		\centering
		\includegraphics[ width=0.3\textwidth]{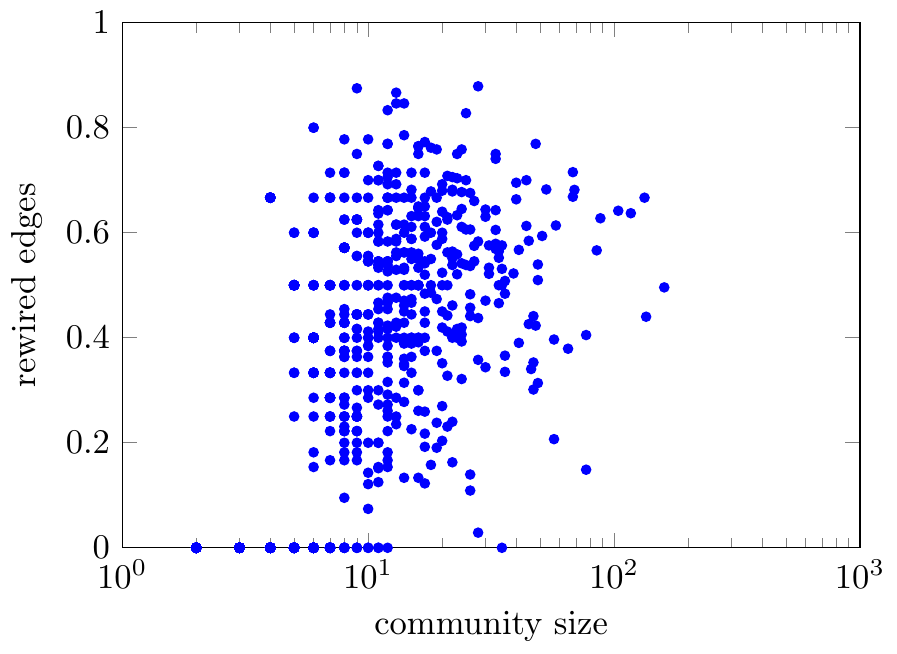}
		
		\label{fig:rewPGP}
	}
	\subfloat[]{
		\centering
		\includegraphics[ width=0.3\textwidth]{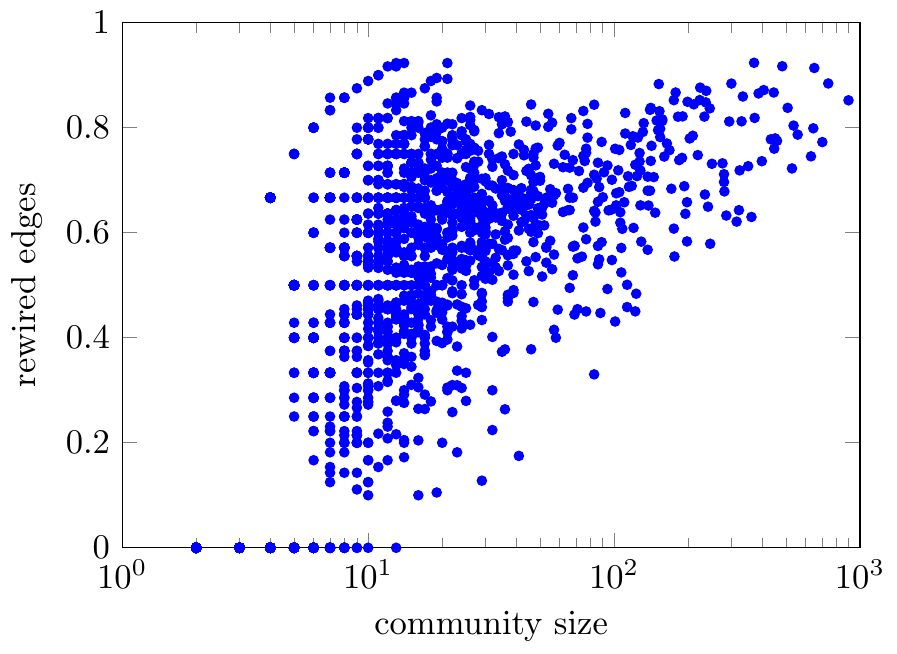}
		
		\label{fig:rewpow}
	}
	\subfloat[]{
		\centering
		\includegraphics[ width=0.3\textwidth]{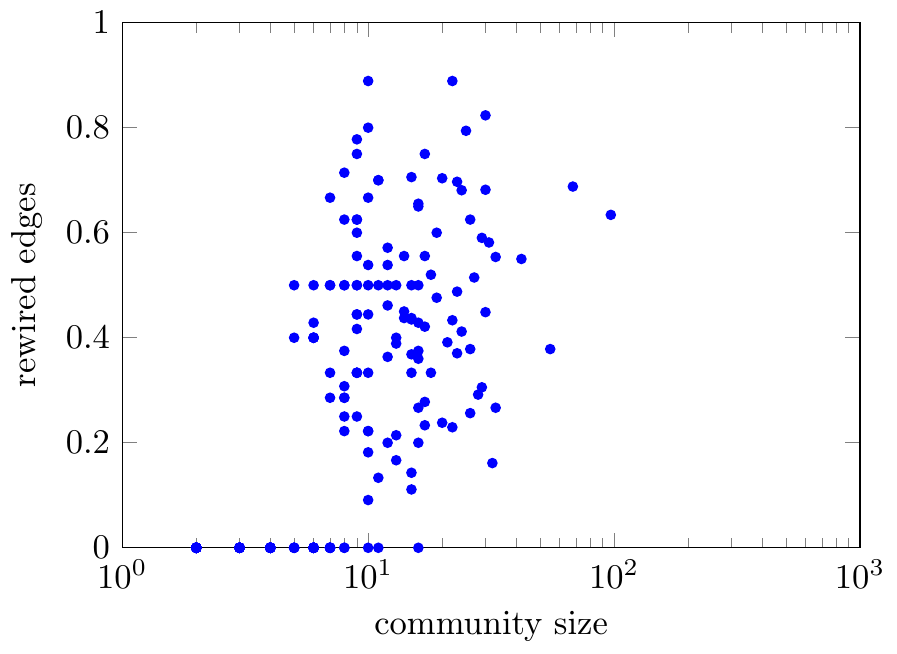}
		
		\label{fig:rewyeast}
	}
	\caption{\textbf{The fraction of rewired edges inside communities for HCM*.} a) Autonomous Systems network b) \textsc{Enron} email network c) Collaboration network in High energy physics d) PGP network e) \textsc{Facebook} friendship network f) yeast network. Every dot corresponds to a community. The fraction of rewired edges is the fraction of edges in the community that are present after randomizing the intra-community edges, but were not present before randomizing.}
	\label{fig:rew}	
\end{figure}

\begin{figure}[!h]
	\centering
	\subfloat[]{
		\centering
		\includegraphics[ width=0.3\textwidth]{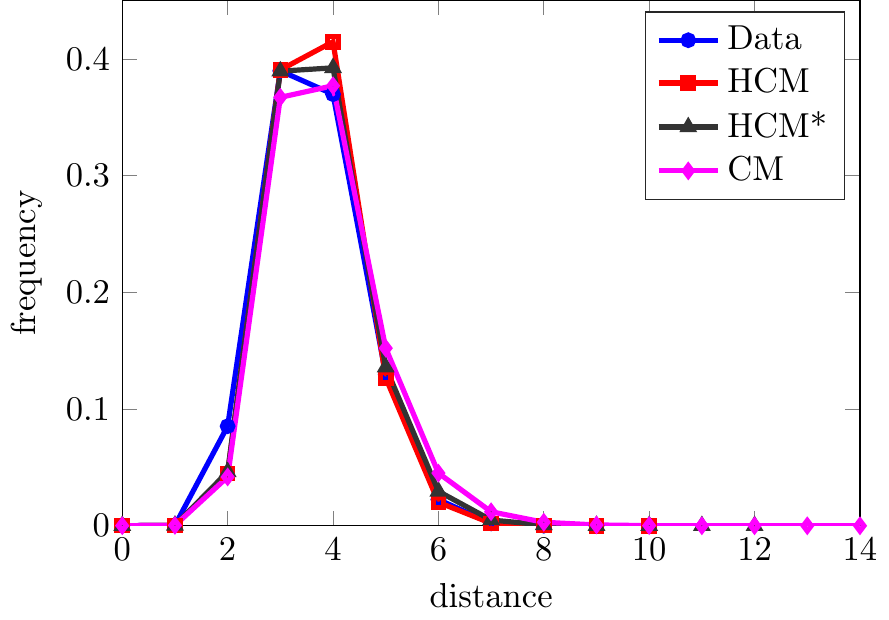}	 
		\label{fig:distas}
	}
	\subfloat[]{
		\centering
		\includegraphics[ width=0.3\textwidth]{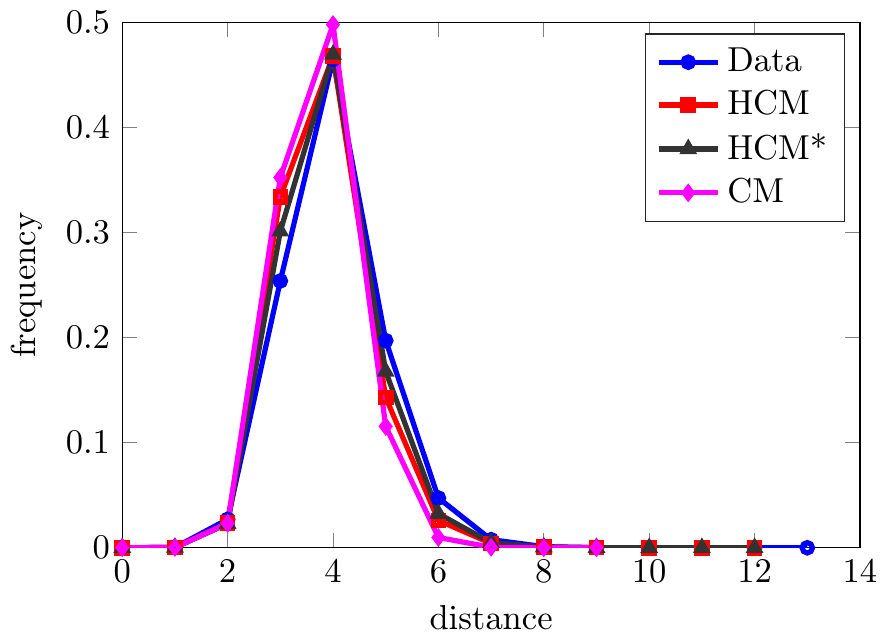}
		
		\label{fig:disten}
	}
	\subfloat[]{
		\centering
		\includegraphics[ width=0.3\textwidth]{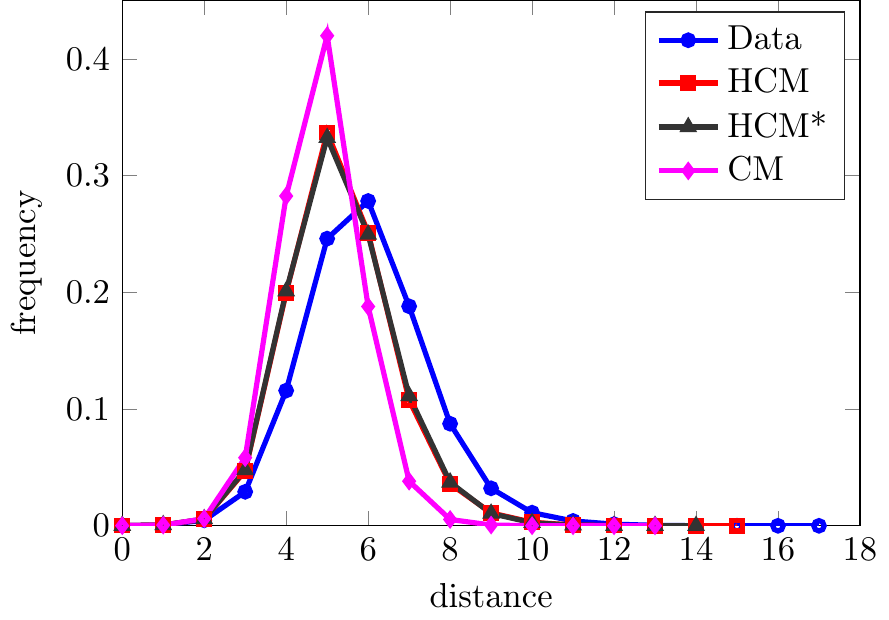}
		
		\label{fig:distHEP}
	}

	\subfloat[]{
		\centering
		\includegraphics[ width=0.3\textwidth]{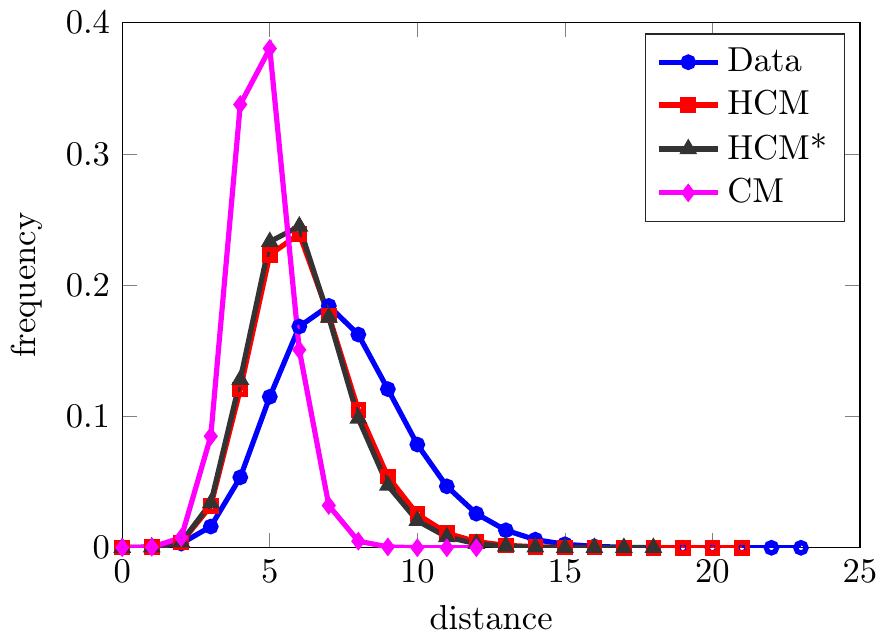}
		
		\label{fig:distPGP}
	}
	\subfloat[]{
		\centering
		\includegraphics[ width=0.3\textwidth]{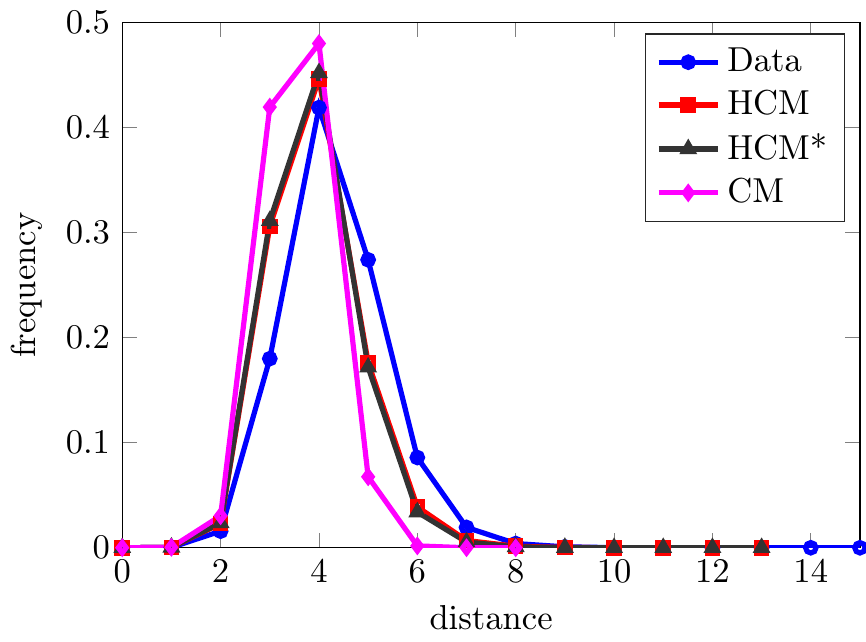}
		
		\label{fig:distpow}
	}
	\subfloat[]{
		\centering
		\includegraphics[ width=0.3\textwidth]{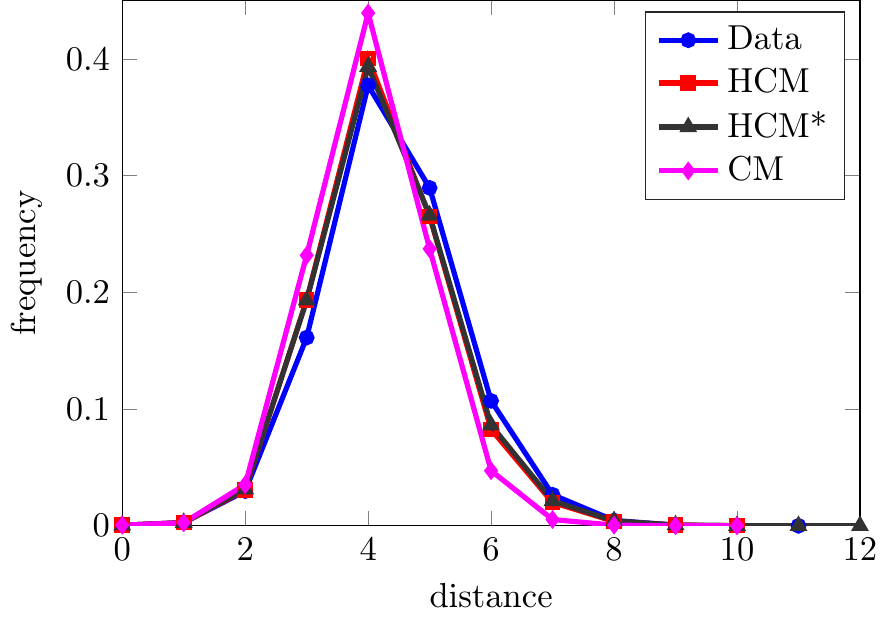}
		
		\label{fig:distyeast}
	}
	\caption{\textbf{Distances in the original network, HCM, HCM* and CM.} a) Autonomous Systems network b) \textsc{Enron} email network c) Collaboration network in High energy physics d) PGP network e) \textsc{Facebook} friendship network f) yeast network. Distances are approximated by sampling 5,000 nodes from the graphs, and calculating all distances between pairs of nodes in the sampled set. The values for HCM, HCM* and CM are the average over 100 generated graphs.}
	\label{fig:dist}	
\end{figure}

%Supplementary Figures~\ref{fig:boot}\--\ref{fig:rec} present the behavior of HCM and HCM* compared to real-world networks under bootstrap percolation, a targeted attack, site percolation and the number of infected and recovered individuals in an SIR epidemic respectively.
Supplementary Table~\ref{tab:rew} shows the fraction of edges of the data sets that are inside communities. HCM fixes all these edges, so one could argue that HCM overfits the data by keeping this fraction of edges fixed. For this reason, we also consider HCM*. 
Supplementary Figure~\ref{fig:rew} shows the fraction of rewired edges in HCM* in communities of size $s$. This is the fraction of edges that are different from the edges in the original community after the rewiring procedure inside communities.  In general, a large fraction of edges is different after randomizing the intra-community edges. The cases where only a few edges were rewired correspond to small communities, where only a small amount of simple random graphs with the same degree distribution exist, or larger communities that are complete graphs, or star-shaped (where only one simple graph with that degree distribution exists). This shows that HCM* creates substantially different graphs than HCM, and is less prone to overfitting the data than HCM.

\section*{Supplementary Tables}
\begin{table}[!h]
	\centering
	\begin{tabular}{lrrrrr}
		\toprule
		& \textbf{data}  & \textbf{HCM}   & \textbf{HCM*}  & \textbf{CM}    & \textbf{HCM (theory)} \\
		\midrule
		\textbf{AS}    & -0.19 & -0.16 & -0.16 & -0.14 & 0.00 \\
		\textbf{Enron} & -0.11 & -0.06 & -0.05 & -0.05 & -0.02 \\
		\textbf{HEP}   & 0.27 & 0.25 & 0.23 & 0.00 & 0.25 \\
		\textbf{PGP}   & 0.24 & 0.26 & 0.26 & -0.01 & 0.26 \\
		\textbf{FB}    & 0.18 & 0.11 & 0.10 & 0.00 & 0.11 \\
		\textbf{yeast} & -0.10 & -0.03 & 0.00 & -0.01 & -0.02 \\
		\bottomrule
	\end{tabular}%
	\caption{Assortativity of HCM, HCM* and CM compared to the real network and theoretical HCM value. The theoretical value is derived in Supplementary Note~\ref{sec:assort}. The values of HCM, HCM* and CM are averages over 500 generated graphs.}
	\label{tab:assort}%
\end{table}%

\begin{table}[htbp]
	\centering
	\begin{tabular}{rrrrrr}
		\toprule
		\textbf{AS} & \textbf{Enron} & \textbf{Hep} & \textbf{PGP} & \textbf{FB} & \textbf{yeast} \\
		\midrule
		0.58 & 0.58 & 0.70 & 0.83 & 0.54 & 0.52 \\
		\bottomrule
	\end{tabular}%
	\caption{The fraction of edges inside communities in the data sets.}
	\label{tab:rew}%
\end{table}%

% Table generated by Excel2LaTeX from sheet 'Density'
\begin{table}[htbp]
	\centering
	\begin{tabular}{rrrrr}
		\toprule
		& $f_{\text{dis}}$  & $N_{\text{dis}}$ & $w_{\text{HCM}}$& $w_{\text{HCM*}}$\\
		\midrule
		\textbf{AS}    & 0.24  & 3.00  & 0.68  & 0.65 \\
		\textbf{Enron} & 0.02  & 3.62  & 0.94  & 0.92 \\
		\textbf{HEP}   & 0.04  & 2.23  & 0.96  & 0.94 \\
		\textbf{PGP}   & 0.17  & 2.54  & 0.97  & 0.91 \\
		\textbf{FB}    & 0.17  & 2.61  & 0.93  & 0.92 \\
		\textbf{yeast} & 0.11  & 2.29  & 0.85  & 0.81 \\
		\bottomrule
	\end{tabular}%
	\caption{Connectedness of HCM* communities and overlap of community structure of generated networks and original networks as defined in~\eqref{eq:sim}.}
	\label{tab:con}%
\end{table}%
\clearpage

\section*{Supplementary Figures}

\end{document}